\documentclass[preprint,12pt]{elsarticle}

\usepackage[T1]{fontenc}
\usepackage[hidelinks]{hyperref}
\usepackage{graphicx}
\usepackage{algorithm}
\usepackage{algpseudocode}
\usepackage{amsmath}
\usepackage{amssymb}
\usepackage{xcolor}
\usepackage{soul}
\usepackage[caption=false]{subfig}
\usepackage{booktabs}
\usepackage{placeins}
\usepackage{natbib}

\sethlcolor{yellow}

\begin{document}

\begin{frontmatter}

\title{Detecting Collusion in Peer Review: Drawing Inspiration from VCG Principle}

\author{Itay Rabinovitz}
\author{Rica Gonen}
\author{Omer Lev}
\author{Asaf Samuel}

\begin{abstract}
The peer-review process, the bedrock of scientific advancement, is increasingly undermined by sophisticated collusion rings that systematically manipulate review outcomes to favor in-group members. While existing detection methods struggle to untangle obfuscated social ties in explicit co-authorship graphs, we introduce a new direction: Exclusion Based Anomaly Detection. Similar to the way VCG auctions work, we formally measure the marginal influence of suspected reviewer groups, exposing their signature even when explicit social graphs are hidden. To apply this at scale without prior knowledge of colluding groups, we introduce the Embedding Based Discovery Framework, which leverages continuous semantic embeddings to isolate latent collusive communities directly from their semantic profile, bypassing the adversarial limitations of explicit network analysis. Unlike traditional heuristic-based approaches, our framework functions as an automated auditor, requiring no prior knowledge of group membership. It achieves this by executing a decoupled search across independent diagnostic algorithms and combining their findings into distinct consensus formations, allowing organizers to dynamically balance detection precision and recall. Evaluating our technique with large-scale datasets (based on ICLR 2021) shows our method's capacity to identify both overt and subtle adversarial tactics with high sensitivity and strict Family-Wise Error Rate (FWER) control, effectively providing conference organizers with a scalable, robust, and privacy-preserving tool to secure the scientific integrity of academic publishing.
\end{abstract}


\end{frontmatter}

\section{Introduction}

The integrity of scientific research relies heavily on rigorous peer review. However, this foundational process is increasingly threatened by organized collusion rings, groups of reviewers who coordinate their evaluations to manipulate the system and maximize the acceptance rate of their members' submissions \cite{Ghosh2024, Rivera2021, Olckers2022, Jecmen2020}. In computer science, where authors frequently serve as reviewers at large competitive conferences, this problem is particularly acute \cite{casey2024, Littman2021, Leyton2024}.

Existing approaches attempt to find strategyproof mechanisms for peer evaluation \cite{KLMP15,ALMRW19,XZSS19,LMTZ23}, but they cannot handle group-strategyproofness considerations. Other algorithmic and mechanism-design methods to tackle collusion rings struggle to untangle obfuscated ties in explicit co-authorship graphs \cite{Jecmen2024, Wu2021, Shah2022, Stelmakh2021}. To address this, we introduce an outcome-based approach: Exclusion Based Anomaly Detection. Similar to the way that VCG mechanisms examine the marginal influence of each agent, we formally measure the marginal influence of suspected colluding groups, exposing their signature even when explicit social graphs are hidden. To achieve this, we formulate three targeted algorithms to capture the distinct behavioral signatures of coordinated manipulation: in-group promotion, out-group suppression, and anomalous localized variance.

To evaluate our proposed algorithms we model a diverse spectrum of collusion strategies, ranging from fully coordinated score inflation to subtle, probabilistic manipulation, and simulate them on data from real-world conference datasets (ICLR 2021 and DPR \cite{kerzendorf2020}). We further examine settings with multiple concurrent collusion rings, demonstrating our algorithms' structural robustness in highly saturated and noisy environments. Results indicate our algorithms effectively separate malicious interference from natural evaluation noise.

While these algorithms successfully evaluate pre-identified suspicious groups, a simple search for covert collusion rings within a massive reviewer population is computationally intractable. To address this, our Embedding Based Discovery Framework leverages the latent semantic structure of reviewer profiles. By projecting reviewers into a continuous embedding space, we isolate core clusters bound by academic commonalities. Our pipeline dynamically refines these cores across the diagnostic algorithms, transforming an intractable combinatorial problem into an efficient discovery process balancing precision and recall.

Unlike traditional heuristic approaches \cite{Shah2022}, our framework functions as an unsupervised detector, requiring no prior knowledge of group membership. This requires some assumptions which are quite common in the peer review domain: First, that colluding agents share an identifiable academic baseline (therefore, fabricated semantic profiles may evade initial clustering). Second, detection requires an honest majority -- a ring that overtakes the reviewer pool dictates the baseline variance, neutralizing anomaly detection. Finally, due to the strict confidentiality of raw peer-review data and the absence of ground-truth collusion labels, our empirical evaluation relies on synthetically injected manipulations.

While we use the terminology of academic peer review, note that these results apply to peer evaluation and selection settings in which agents may attempt to form collusion rings to influence the outcome of the mechanism, and in which our assumptions above hold.
Recognizing these boundaries, our main contributions are as follows:

\begin{description}
 \item [Marginal-Outcome Detection Approach] A formal, Exclusion Based framework evaluating the collusive power of reviewer groups.
 \item [Targeted Detection Algorithms] Three specialized algorithms quantifying in-group promotion, out-group suppression, and scoring variance.
 \item [Embedding Based Discovery Framework] A scalable, semantic-driven mechanism to automatically isolate potential collusive communities.
 \item [Simulation-Based Evaluation] Using simulations on real-world datasets (augmented with synthetic collusion, which is, of course, unknowable from the raw data).
 \item [Robustness Under Concurrent Collusion] Maintaining high detection sensitivity and discovery precision in environments with multiple collusion rings.
\end{description}

\section{Related Work}
Concerns about the integrity of the peer-review process, particularly in computer science conferences, have intensified in recent years. The literature broadly categorizes anti-collusion efforts into two main approaches: \textit{prevention} and \textit{detection}. Various strategies have been proposed, including strategyproof mechanisms, algorithmic detection methods, and institutional policy reforms.

Early research focused primarily on mitigating individual reviewer bias and manipulation. Approaches such as double-blind review and reviewer blacklisting were introduced to reduce bias and discourage dishonest behavior by individual reviewers \cite{Thurner2011,DAndrea2017,Jecmen2020,Zhang2022}. However, more recent studies highlight the emergence of group-based collusion, in which coordinated reviewers strategically exploit the system to promote one another’s submissions \cite{Littman2021,Jecmen2020,Olckers2022,Leyton2024}.

To address this threat, several strategyproof mechanisms have been proposed. While early distributed review proposals were highly susceptible to strategic manipulation~\cite{Merrifield2003}, later approaches built on foundational mechanism design principles for peer selection~\cite{Alon2011}. For instance, \citet{Aziz2019} and \citet{Lev2023} designed randomized, strategyproof algorithms (such as \textsc{PeerNomination}) that handle both noisy assessments and strategic manipulation by decoupling a reviewer's own success from the evaluations they provide to others. Although theoretically sound, such approaches face practical challenges, including fairness constraints and the difficulty of identifying undeclared conflicts of interest. \citet{Jecmen2024} introduced a complementary detection strategy based on bidding data; however, their approach exhibited low precision in identifying collusion in practice.

Other studies have explored machine learning and anomaly detection techniques to audit reviewer behavior. \citet{Stelmakh2021} developed statistical models specifically designed to catch strategic manipulation and malicious penalties in peer assessments. \citet{Wu2021} proposed a predictive model that estimates a reviewer's bidding probability based on their publication history. Similarly, \citet{Shah2022} introduced heuristic indicators of collusion by identifying anomalously high reviewer-author scoring patterns.

In parallel, institutional interventions have become an important complement to algorithmic approaches. Conferences have introduced sanctions for collusive behavior, including revoking future reviewing privileges. Program chairs increasingly audit conflict-of-interest (COI) declarations and investigate potential inconsistencies. Automated tools that leverage co-authorship graphs and collaboration networks have also been deployed to identify undeclared conflicts of interest.

Despite these advances, collusion remains a persistent and elusive challenge. Colluding reviewers often operate subtly, adapt quickly, and exploit structural weaknesses in existing systems. As a result, no single approach, whether mechanism-based, algorithmic, or institutional, has proven sufficient on its own, underscoring the need for complementary methodologies that can reliably detect coordinated manipulation in peer review.

Concurrently, the application of Natural Language Processing (NLP) has transformed reviewer assignment systems. Modern conference platforms increasingly utilize semantic embeddings to compute manuscript reviewer similarity scores, optimizing assignments based on textual research content rather than solely relying on self-reported expertise. While these semantic tools were primarily designed for operational efficiency \cite{kerzendorf2020}, their latent capacity for structural auditing and anomaly detection remains largely unexplored in the context of academic collusion.

\section{Theoretical Framework and Collusion Dynamics}
In this section, we formally define the problem of detecting collusive behavior in a peer-review system.

\subsection{Preliminaries}
We consider a standard peer-review setting in which reviewers are assigned submissions and evaluate them by assigning numeric scores from a predefined domain. Let $\mathcal{R}$ denote the set of $m$ reviewers, and let $\mathcal{S}$ denote the set of $n$ submissions. The authorship matrix $A \in \{0,1\}^{m \times n}$ indicates if $r_i$ is an author of $s_j$. A single reviewer may author multiple submissions, and submissions may have multiple authors.

Let $\mathcal{V}$ denote the domain of valid review scores and $0 \notin \mathcal{V}$ (e.g., $\mathcal{V} = \{1, 2, \dots, 10\}$ or $\mathcal{V} = [1.0, 5.0]$). Review scores are represented by the evaluation matrix $E \in (\mathcal{V} \cup \{0\})^{m \times n}$, where $E_{i,j}$ denotes the score\footnote{While some mechanisms use approval or rankings, we shall assume scores, as they can be translated into top-$k$ approval or to rankings.} given by reviewer $r_i$ to submission $s_j$. Let $\mathcal{R}_j \subseteq \mathcal{R}$ denote the specific reviewers assigned to evaluate $s_j$. We adopt the convention that $E_{i,j} = 0$ indicates reviewer $r_i$ was not assigned to $s_j$. After all reviews are submitted, some peer selection mechanism chooses $w$ submissions to be accepted\footnote{Many conferences simply rank paper grade average, but more elaborate mechanisms have been suggested.}. Let $\hat{\mathcal{S}} \subseteq \mathcal{S}$ denote the set of accepted submissions. To investigate potential collusion, let $G \subseteq \mathcal{R}$ denote a specific candidate colluding group suspected of manipulating outcomes.

\subsection{VCG-Inspired Collusion Dynamics}
Our objective is to identify and characterize collusive behavior within the peer-review process using principles derived from mechanism design. Specifically, our approach of excluding a suspected coalition and measuring its counterfactual effect on the outcome is conceptually rooted in the Vickrey-Clarke-Groves (VCG) marginal contribution principle. In classical game theory, a participant's true influence is quantified by comparing the global outcome with and without their presence. By applying this principle to peer review, we can objectively measure whether a subset of reviewers is coordinating to disproportionately influence acceptance decisions for mutual benefit.

Colluding reviewers employ various strategies, all ultimately aimed at increasing the acceptance rate of papers submitted by their group members. To evade detection by conference organizers, these coalitions utilize sophisticated methods designed to minimize the statistical anomalies caused by their manipulated scores. Consequently, there is an inherent trade-off regarding group size: while a larger ring has a greater capacity to alter the overall conference outcomes, its expanded footprint makes it significantly more difficult to hide.

Ultimately, regardless of the specific evasion strategy utilized, the fundamental conflict between coordinated manipulation and independent, objective evaluations inevitably introduces structural anomalies. Submissions targeted by a collusion ring, whether for promotion or suppression, often receive a mix of highly biased scores from colliders and honest, quality-based evaluations from independent reviewers. This clash typically manifests as an anomalously high variance or a bimodal distribution in the scores of the affected submissions, leaving a distinct, measurable footprint that our framework is designed to detect.

\section{Statistical Detection Algorithms}
In the Vickrey-Clarke-Groves (VCG) mechanism, each agent is removed from the setting, and the outcome without its input is calculated. We use this method in determining whether a suspected subset of reviewers exhibits collusive behavior by removing their reviews and measuring the resulting counterfactual changes. The statistical significance of all three proposed metrics ($\delta_G, \Delta_G, \Sigma^2_G$) is established via a unified permutation testing framework, detailed below.

All these algorithms are agnostic regarding the actual peer-selection mechanism used, and allow measuring the effects of collusion rings on any mechanism used for this purpose.

\subsection{Algorithm 1: In-Group Promotion Influence}
\label{subsec:Algorithm 1}
To evaluate the impact of a suspected colluding group $G \subseteq \mathcal{R}$ on the acceptance of its own submissions, we measure the counterfactual change in acceptance outcomes induced by removing their evaluations.

Let $\bar{E_{G}}$ denote the modified exclusion matrix, representing the review environment without the active participation of $G$. Let $\hat{\mathcal{S}}_E$ and $\hat{\mathcal{S}}_{\bar{E_{G}}}$ denote the sets of the top $w$ submissions accepted based on the full review matrix $E$ and the modified matrix $\bar{E_{G}}$, respectively.\footnote{Throughout this work, if the exclusion of $G$ leaves a submission with no remaining reviews, that submission is entirely omitted from all subsequent score aggregations and acceptance considerations.} Let $\mathcal{S}_{G}$ denote the set of submissions authored or co-authored by members of $G$.

We measure the influence of $G$ through the change in the proportion of accepted in-group submissions relative to the total acceptance capacity $w$:
\small
\[
\delta_{G} = \frac{|\hat{\mathcal{S}}_E \cap \mathcal{S}_{G}| - |\hat{\mathcal{S}}_{\bar{E_{G}}} \cap \mathcal{S}_{G}|}{w} \times 100
\]
\normalsize

\begin{algorithm}[!ht]
\begin{small}
    \caption{\label{alg:algo1}Influence on In-Group Submissions}
    \textbf{Input:} Reviewer set $\mathcal{R}$, submission set $\mathcal{S}$, review matrix $E$, authorship matrix $A$, acceptance quota $w$, suspected group $G$, number of trials $T$, significance level $\gamma$.
    
    \textbf{Output:} A tuple $(\textit{IsCollusive}, \delta_G)$, where \textit{IsCollusive} is \texttt{True} if $G$ exhibits collusive behavior, and $\delta_G$ represents the percentage-point drop in in-group acceptance.
    \begin{algorithmic}[1]
        \State Compute $\hat{\mathcal{S}}_E$ from the full review matrix $E$ as the top $w$ submissions.
        \State Construct $\bar{E_{G}}$:
        $$
        (\bar{E_{G}})_{i,j} =
        \begin{cases}
            0, & \text{if } r_i \in G, \\
            E_{i,j}, & \text{otherwise.}
        \end{cases}
        $$
        \State Compute $\hat{\mathcal{S}}_{\bar{E_{G}}}$ from the review matrix $\bar{E_{G}}$ as the top $w$ submissions.
        \State Let $\mathcal{S}_{G} \gets \{ s_j \in \mathcal{S} \mid \exists\, r_i \in G,\ A_{i,j} = 1 \}$.
        \State Compute:
        \[
        \delta_{G} = \frac{|\hat{\mathcal{S}}_E \cap \mathcal{S}_{G}| - |\hat{\mathcal{S}}_{\bar{E_{G}}} \cap \mathcal{S}_{G}|}{w} \times 100\]
        \For{$t = 1$ to $T$}
            \State Sample $G^{(t)} \subseteq \mathcal{R}$ with $|G^{(t)}| = |G|$.
            \State Compute $\delta^{(t)}$ by repeating Steps 2-5 with $G^{(t)}$.
        \EndFor
        \State Compute permutation $p$-value: \quad $p \gets \dfrac{1 + \sum_{t=1}^T \mathbb{I}\!\left[\, \delta^{(t)} \ge \delta_{G} \,\right]}{T+1}$
        \State \Return $(\mathbb{I}[p < \gamma], \delta_G)$
    \end{algorithmic}
\end{small}
\end{algorithm}

\subsection{Algorithm 2: Out-Group Suppression Influence}
We next assess whether $G$ exerts a suppressive influence on submissions not affiliated with its members. We define the set of competing submissions as:
\small
\[
\mathcal{S} \setminus \mathcal{S}_{G}
= \left\{ s_j \in \mathcal{S} \ \middle| \ \forall \, r_i \in G,\ A_{i,j} = 0 \right\}.
\]
\normalsize
For any review matrix $E$ and submission $s_j$, let:
\small
\[
\operatorname{AvgRating}_E(s_j) = \frac{1}{|\{r_i \in \mathcal{R}_j \mid E_{i,j} > 0\}|} \cdot \sum_{r_i \in \mathcal{R}_j} E_{i,j}
\]
\normalsize
denote its average review score under the full evaluation matrix $E$.
The external influence of $G$ is then defined as:
\small
\[
\Delta_{G}
=
\sum_{s_j \in \mathcal{S} \setminus \mathcal{S}_{G}}
\left(
\operatorname{AvgRating}_{\bar{E_{G}}}(s_j)
-
\operatorname{AvgRating}_E(s_j)
\right).
\]
\normalsize

\begin{algorithm}[!ht]
\begin{small}
    \caption{\label{alg:algo2}Influence on Competing Submissions}
    \textbf{Input:} Reviewer set $\mathcal{R}$, submission set $\mathcal{S}$, review matrix $E$, authorship matrix $A$, suspected group $G$, number of trials $T$, significance level $\gamma$.
    
    \textbf{Output:} A tuple $(\textit{IsCollusive}, \Delta_G)$, where \textit{IsCollusive} is \texttt{True} if $G$ exhibits collusive behavior; otherwise \texttt{False}.
    \begin{algorithmic}[1]
        \State Construct $\bar{E_{G}}$ by removing reviews written by members of $G$
        \State Let $\mathcal{S}~\setminus~\mathcal{S}_{G} = \{ s_j \in \mathcal{S} \mid \forall \, r_i \in G,\ A_{i,j} = 0 \}$
        \State Compute:
        \[
        \Delta_{G} =
        \sum_{s_j \in \mathcal{S} \setminus \mathcal{S}_{G}}
        \left(
        \operatorname{AvgRating}_{\bar{E_{G}}}(s_j)
        -
        \operatorname{AvgRating}_E(s_j)
        \right)
        \]
        \For{$t = 1$ to $T$}
            \State Sample $G^{(t)} \subseteq \mathcal{R}$ with $|G^{(t)}| = |G|$
            \State Compute $\Delta^{(t)}$ by repeating Steps 1-3 with $G^{(t)}$
        \EndFor
        \State Compute permutation $p$-value: \quad $p \gets \dfrac{1 + \sum_{t=1}^T \mathbb{I}\!\left[\, \Delta^{(t)} \ge \Delta_{G} \,\right]}{T+1}$
        \State \Return $(\mathbb{I}[p < \gamma], \Delta_G)$
\end{algorithmic}
\end{small}
\end{algorithm}

\subsection{Algorithm 3: Scoring Variance Analysis}
\label{subsec:Algorithm 3}
Targeted submissions often receive a bimodal mix of biased scores from colliders and objective scores from honest reviewers. To formally capture this discrepancy, we compute the sample variance for each in-group submission $s_j \in \mathcal{S}_{G, \ge 2}$ (where $\mathcal{S}_{G, \ge 2} = \{s_j \in \mathcal{S}_G \mid |\mathcal{R}_j| \ge 2\}$) as $\operatorname{Var}(s_j) = \frac{1}{|\mathcal{R}_j| - 1} \sum_{r_i \in \mathcal{R}_j} \left( E_{i,j} - \operatorname{AvgRating}_E(s_j) \right)^2$. The structural inconsistency is quantified by the mean variance across all group-authored submissions:
\small
\[
\Sigma^2_{G} = \frac{1}{|\mathcal{S}_{G, \ge 2}|} \sum_{s_j \in \mathcal{S}_{G, \ge 2}} \operatorname{Var}(s_j).
\]
\normalsize

\begin{algorithm}[!ht]
\begin{small}
    \caption{\label{alg:algo3}Detection of Scoring Inconsistency (Variance)}
    \textbf{Input:} Reviewer set $\mathcal{R}$, submission set $\mathcal{S}$, review matrix $E$, authorship matrix $A$, suspected group $G$, number of trials $T$, significance level $\gamma$.
    
    \textbf{Output:} A tuple $(\textit{IsCollusive}, \Sigma^2_G)$, where \textit{IsCollusive} is \texttt{True} if $G$ exhibits collusive behavior; otherwise \texttt{False}.
    \begin{algorithmic}[1]
        \State Let \( \mathcal{S}_{G} \gets \{ s_j \in \mathcal{S} \mid \exists\, r_i \in G,\ A_{i,j} = 1 \} \)
        \State For each submission $s_j \in \mathcal{S}_{G}$ with $|\mathcal{R}_j| \ge 2$, compute the sample variance of its ratings:
        \[
        \operatorname{Var}(s_j) = \frac{1}{|\mathcal{R}_j| - 1} \sum_{r_i \in \mathcal{R}_j} \left( E_{i,j} - \operatorname{AvgRating}_E(s_j) \right)^2
        \]
        \State Compute the mean variance across all group-authored submissions:
        \[
        \Sigma^2_{G} = \frac{1}{|\mathcal{S}_{G, \ge 2}|} \sum_{s_j \in \mathcal{S}_{G, \ge 2}} \operatorname{Var}(s_j).
        \]
        \For{$t = 1$ to $T$}
            \State Sample $G^{(t)} \subseteq \mathcal{R}$ with $|G^{(t)}| = |G|$
            \State Compute $\Sigma^{2(t)}$ by repeating Steps 1-3 with $G^{(t)}$
        \EndFor
        \State Compute permutation $p$-value: \quad $p \gets \dfrac{1 + \sum_{t=1}^T \mathbb{I}\!\left[\, \Sigma^{2(t)} \ge \Sigma^2_{G} \,\right]}{T+1}$
        \State \Return $(\mathbb{I}[p < \gamma], \Sigma^2_G)$
\end{algorithmic}
\end{small}
\end{algorithm}

\subsection{Unified Statistical Significance and FPR Control}
\label{subsec:significance_testing}

To assess the statistical significance of an observed metric $\mathcal{M}_G \in \{\delta_G, \Delta_G, \Sigma^2_G\}$ for a suspected group $G$, we deploy a unified permutation test. We construct an empirical null distribution by repeatedly sampling $T$ random reviewer groups $G^{(t)} \subseteq \mathcal{R}$ of size $|G|$ and computing their baseline measures $\{\mathcal{M}^{(t)}\}_{t=1}^T$. The permutation $p$-value is calculated as:
\small
\[
p = \frac{1 + \sum_{t=1}^T \mathbb{I}\!\left[ \mathcal{M}^{(t)} \geq \mathcal{M}_{G} \right]}{1 + T}
\]
\normalsize
where $\mathbb{I}[\cdot]$ is the indicator function. We reject the null hypothesis of non-collusive behavior if $p < \gamma$ (e.g., $\gamma=0.01$).

\textbf{Empirical Null and FPR Control:} 
Crucially, this null distribution is generated directly from the observed, potentially contaminated, review matrix $E$. Thus, a flagged group's influence must uniquely stand out against the actual background noise of the venue, not merely an idealized clean state. This non-parametric procedure bounds the False Positive Rate (FPR) at the chosen $\gamma$, explicitly protecting innocent collaborations.
To demonstrate this exact error boundary mathematically, let $Z^*$ denote the observed collusion score for a suspected group, and $\{Z_1, \dots, Z_T\}$ be the baseline scores under the null hypothesis $H_0$. Because the permutation-based $p$-value is a valid test statistic under $H_0$, the probability of incorrectly rejecting the null hypothesis is mathematically bounded by $\gamma$. Therefore, the false positive rate satisfies:
$$
FPR = Pr(p < \gamma \mid H_0) \le \gamma
$$

Thus, selecting $\gamma$ directly controls the Type I error rate without requiring parametric assumptions about the distribution of the score.

\section{Experimental Setup and Simulation Methodology}
To evaluate the effectiveness of our proposed collusion detection framework, we conducted a series of experiments using real-world peer-review data augmented with controlled synthetic manipulations. The objectives of our evaluation were fourfold: (1) assess the sensitivity of our detection algorithms under various collusion scenarios, (2) analyze how the size and strategy of colluding groups affect detectability, (3) benchmark the results against randomized baselines, and (4) estimate the FPR and TPR of each algorithm.

\subsection{Conference Datasets (ICLR \& DPR)} \label{Conference Datasets}
We used two complementary datasets to capture different scales and characteristics of the peer review process: publicly available data from the \emph{ICLR 2021} conference (via the OpenReview API) and \emph{DeepThought DPR Dataset I} from \citet{kerzendorf2020}. The ICLR dataset represents a large, high-stakes conference environment, while the DPR dataset reflects a smaller and more controlled review setting. Evaluating our algorithms on both enables us to study their robustness across different conference scales and reviewer dynamics.

\paragraph{ICLR Dataset:}
The ICLR 2021 dataset contains metadata for 2,594 submissions and 10,022 review reports, along with final acceptance decisions (859 accepted, 1,735 rejected), yielding an empirical acceptance rate of approximately 33\%.

Because the identities and assignments of the reviewers are anonymized in this dataset, we employed a simulation-based reconstruction of the review matrix. Each submission ID was treated as a proxy for a unique reviewer ID, reflecting the standard assumption that every author also serves as a reviewer. Review assignments were then generated randomly under two constraints:
\begin{itemize}
    \item No reviewer was assigned to their own submission.
    \item No reviewer reviews the same submission more then one time.
\end{itemize}

Each reviewer was assigned to review 3–4 submissions, producing a review matrix suitable for testing various collusion scenarios.

\paragraph{DPR Dataset:}
The DeepThought Peer Review (DPR) dataset contains metadata for 172 submissions, 136 reviewers, and $1,025$ reviews. Unlike ICLR data, this dataset includes known reviewer identities, assignments, and authorship information. Because the dataset originates from a controlled experimental setting - with voluntary participation and no implications for real conference outcomes - it contains no collusion by design. As a result, we can use it directly, without reconstruction, as a baseline representing a fully honest review environment.

\subsection{Modeling Collusion Strategies}
\label{subsec:collusive_behavior}
To evaluate our detection framework under diverse adversarial conditions, we simulate collusive manipulation by randomly selecting a subset of reviewers to form a colluding group and altering their assigned review scores according to predefined behavioral models. 
Each model captures a distinct strategic pattern observed or hypothesized in real-world peer review. All other reviewers behave honestly, assigning scores based on independent and unbiased assessments.

The framework dynamically adapts the manipulation magnitude based on the conference's scoring domain, where $S_{max}$ denotes the maximum possible score. We model a spectrum of adversarial behaviors, implementing the following collusion types:
\begin{enumerate}
    \item \textbf{No Collusion (Type 1):} The baseline control scenario where all reviews reflect independent, unbiased evaluations.
    
    \item \textbf{Fully Coordinated Collusion (Type 2):} Represents an aggressive strategy where colluders assign maximum scores (e.g., $s \in [0.8, 1.0] \cdot S_{max}$) to in-group submissions and severe penalties (e.g., $s \in [0.3, 0.4] \cdot S_{max}$) to out-group competitors.
    
    \item \textbf{In-Group Promotion Only (Type 3):} A unilateral tactic which exclusively inflates in-group papers (e.g., $s \in [0.8, 1.0] \cdot S_{max}$) while evaluating out-group papers honestly.
    
    \item \textbf{Out-Group Suppression Only (Type 4):} A unilateral tactic which solely penalizes external competitors (e.g., $s \in [0.3, 0.4] \cdot S_{max}$) while evaluating in-group papers honestly.
    
    \item \textbf{Mild Coordinated Bias 1 (Type 5):} An evasion tactic applying a fixed $\pm 0.1 \cdot S_{max}$ shift (positive to in-group, negative to out-group) to evade standard detection.
    
    \item \textbf{Mild Coordinated Bias 2 (Type 6):} A stronger variant of this evasion tactic applying a $\pm 0.2 \cdot S_{max}$ shift to maximize impact without triggering extreme-score filters.
    
    \item \textbf{Probabilistic Collusion (Type 7):} Models a stochastic approach where reviewers execute Type 2 collusion as an independent Bernoulli trial with probability $\phi = 0.5$, scoring honestly otherwise.
    
    \item \textbf{Borderline Manipulation (Type 8):} Represents a strategic ``swing'' model that minimizes conspicuous deviations by exclusively targeting papers near the assumed acceptance threshold, representing the critical boundary where minor score shifts dictate final decisions. In this scenario, mediocre in-group submissions ($[0.4, 0.6] \cdot S_{max}$) are elevated to $0.8 \cdot S_{max}$, while strong out-group submissions ($[0.6, 0.8] \cdot S_{max}$) are downgraded to $0.4 \cdot S_{max}$. Clear accepts or rejects (e.g., out-group $> 0.8 \cdot S_{max}$) are scored honestly, as penalizing them yields negligible utility while drastically increasing detection risk.
\end{enumerate}

In all scenarios, the baseline behavior of honest reviewers remains unaltered. Crucially, the inclusion of the advanced adversarial models (Types 7 and 8) highlights a fundamental limitation of mean-based detection mechanisms. Sophisticated colluders employing these subtle tactics actively minimize their average score footprint to bypass Algorithms 1 and 2. Nevertheless, this strategic restraint inevitably generates localized statistical anomalies and high-variance disagreements with honest peers. Recognizing this distinct signature motivates the development of Algorithm 3, which is specifically designed to detect such variance-driven manipulation.

\subsection{Evaluation Procedure}
To emulate realistic dynamics where colluding reviewers position themselves to evaluate in-group work (e.g., via bidding behavior or topic similarity exploitation), we introduce a positive assignment bias into our simulation. Specifically, the baseline probability $q$ of a reviewer from group $G$ being assigned to a co-conspirator's submission is multiplied by a factor of $\beta = 2.5$.

We define the adjusted probability as $q' = \beta q$, which increases the likelihood of within-group review assignments. For example:
\begin{itemize}
    \item For $\lvert G \rvert = 20$ and $n = 2,594$, where $q = 0.0077$, the biased probability becomes $0.01927$.    
    \item For $\lvert G \rvert = 140$ and $n = 2,594$, where $q = 0.0539$, the biased probability becomes $0.13492$.    
\end{itemize}

\paragraph{Choice of $\beta$}
We explicitly calibrated $\beta = 2.5$ as a realistic upper bound for stealthy collusion, intentionally modest but high enough to make an impact. Prior work and anecdotal evidence suggest that colluders rarely achieve complete control over the assignment process but can substantially increase their chances of being assigned to in-group submissions through bidding behavior, topic similarity exploitation, or strategic conflict declarations. A factor of 2.5 represents a moderate but meaningful deviation from randomness: it increases the likelihood of in-group assignments sufficiently to reveal detectable collusive effects while avoiding implausibly strong bias that would make detection trivial. Empirically, we found that lower values (e.g., $\beta = 1.5$) often yielded minimal observable manipulation effects, while higher values (e.g., $\beta \ge 4$) produced unrealistic assignment patterns inconsistent with documented cases of collusion.

To assess robustness and evaluate the framework under extreme stress, we systematically varied the colluding group size $|G|$ ($2$ to $150$ members in ICLR; $2$ to $70$ in DPR), executing 50 independent runs per configuration. Each simulation run follows a strict pipeline:

\begin{enumerate}
    \item Randomly sample $G \subset \mathcal{R}$.
    \item Construct the assignment matrix utilizing the aforementioned $\beta=2.5$ in-group bias.
    \item Inject the chosen collusion strategy into $G$'s evaluations.
    \item Recalculate global acceptance decisions based on the manipulated averages.
    \item Execute Algorithms 1--3. These algorithms autonomously generate $T=200$ randomized baseline groups to compute the empirical null distributions, evaluating the statistical significance ($p$-value) and returning the final binary classification.
\end{enumerate}

\FloatBarrier

\section{Controlled Simulation Results}
\label{sec:simulation_results}

This section presents the empirical evaluation of our proposed framework based on these simulated environments. We first establish the systemic impact of collusion on review metrics (Effect Size) and subsequently evaluate the framework's detection sensitivity (TPR), structural soundness (FPR), and raw signal separability. To ensure the reproducibility of our results, the source code used for all experiments will be made publicly available.

\subsection{Systemic Impact: Effect Size Dynamics}
\label{subsec:effect_size_dynamics}

Before assessing detection performance, we must validate the systemic impact of the injected manipulations. Broadly, collusion inherently distorts the evaluation baseline, with the magnitude of this footprint dictated by the coalition's size ($|G|$) and their chosen strategy (\textbf{Types 2--8}). While aggressive, \textbf{Fully Coordinated} attacks overtly skew global acceptance thresholds, sophisticated adversarial tactics specifically minimize this raw footprint to evade traditional mean-based filters.

Figures~\ref{fig:all_effect_sizes} (a-f) explicitly map this scaling impact across all algorithms for both the ICLR and DPR datasets, demonstrating how our algorithms capture orthogonal dimensions of this interference. As predicted, \textbf{Type 2 (Fully Coordinated)} collusion produces the most substantial distortions across all metrics. In Algorithm~\ref{alg:algo1}, we observe a significant drop in acceptance rates for colluding groups as their internal support is removed, while Algorithm~\ref{alg:algo2} tracks the near-linear accumulation of massive cumulative damage inflicted on external competitors.

Crucially, the \textbf{Type 1 (Control)} group remains strictly at the zero baseline for promotional and suppressive deltas, while stably reflecting the natural baseline scoring variance of the venue in Algorithm~\ref{alg:algo3}. This confirms that our metrics strictly isolate genuine manipulation from natural evaluation noise.

\subsection{Detection Sensitivity: True Positive Rate (TPR)}
\label{subsec:tpr_analysis}

Having established that collusion creates a distinct evidentiary footprint, we evaluate how effectively our algorithms reject the null hypothesis.

\subsubsection{ICLR Performance:}
As shown in Figure~\ref{fig:tpr_master_all} (c), Algorithm~\ref{alg:algo2} exhibits the highest baseline sensitivity, reaching near-perfect detection for suppression tactics at compact sizes ($|G|\approx 20$). Algorithm~\ref{alg:algo1} (Figure~\ref{fig:tpr_master_all} a) additionally identifies coordination as groups reach a critical mass ($|G|\approx 50$), where their internal inflation successfully overcomes discrete venue noise. Finally, Algorithm~\ref{alg:algo3} (Figure~\ref{fig:tpr_master_all} e) demonstrates proficiency in detecting overt manipulation strategies, specifically \textbf{Fully Coordinated} collusion and \textbf{In-Group Promotion Only}, once the colluding coalition grows sufficiently large ($|G|\ge 80$), but fails in more subtle collusion types, even at the maximum group size tested.

Crucially, the true strength of the proposed methodology emerges when evaluating the combined maximum detection power, Figure~\ref{fig:tpr_master_all} (g). The ensemble approach does more than cover the inherent blind spots of individual algorithms; it fundamentally accelerates the detection threshold. While standalone metrics generally require a critical mass of $|G|\approx 50$ to secure near-perfect true positive rates, the unified framework synergizes orthogonal multi-dimensional signals to achieve comparable detection certainty at significantly smaller coalition sizes (e.g., $|G|\approx 20$). This proves that aggregating weak, localized anomalies provides a highly sensitive early-warning mechanism against emerging collusion rings.

\subsubsection{DPR Performance:}
\begin{figure*}[!ht]
    \subfloat[]{
        \includegraphics[width=0.46\textwidth]{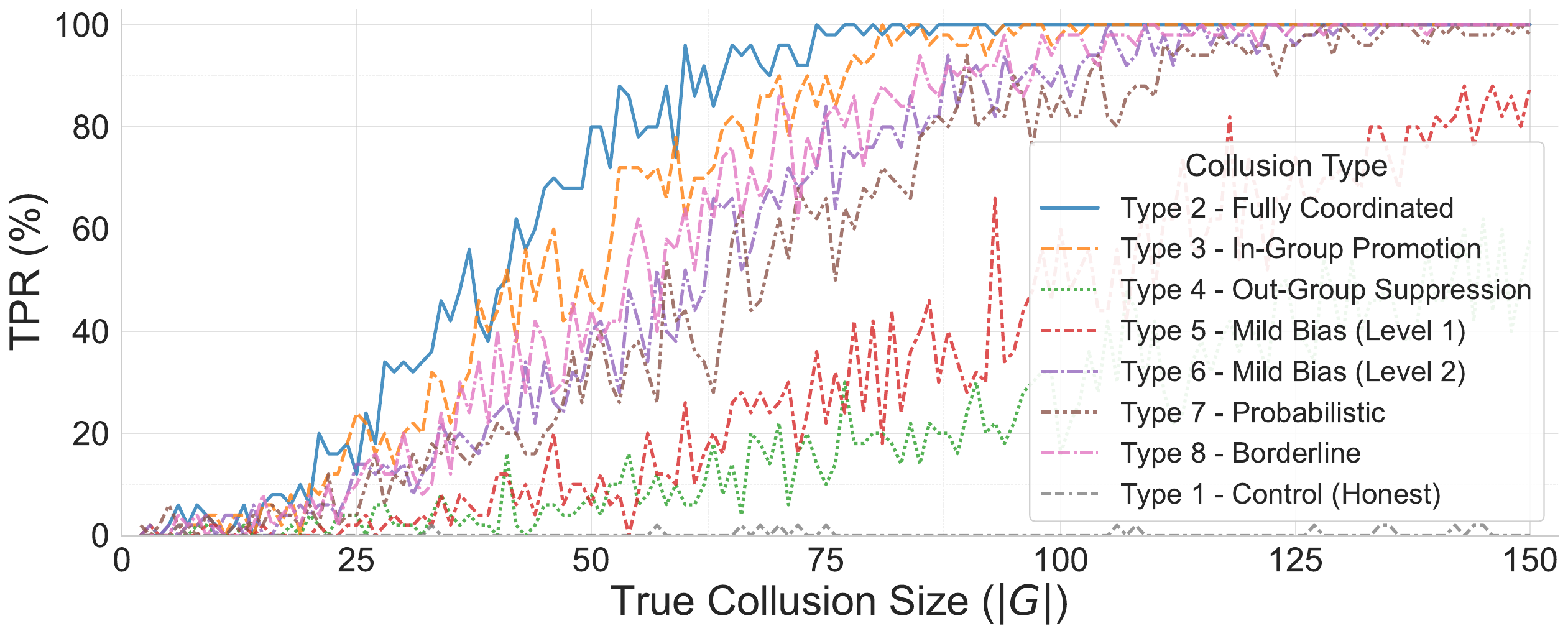}
    }\hfill
    \subfloat[]{
        \includegraphics[width=0.46\textwidth]{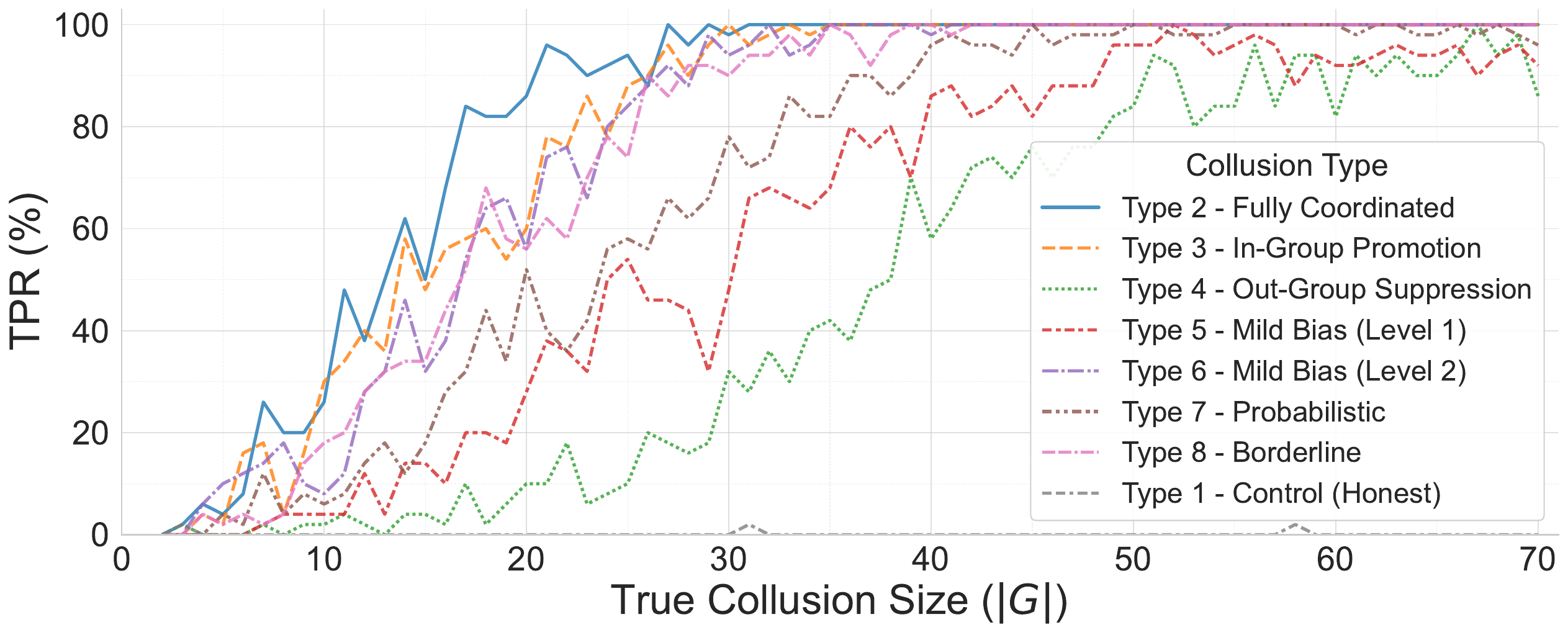}
    }
    
    \subfloat[]{
        \includegraphics[width=0.46\textwidth]{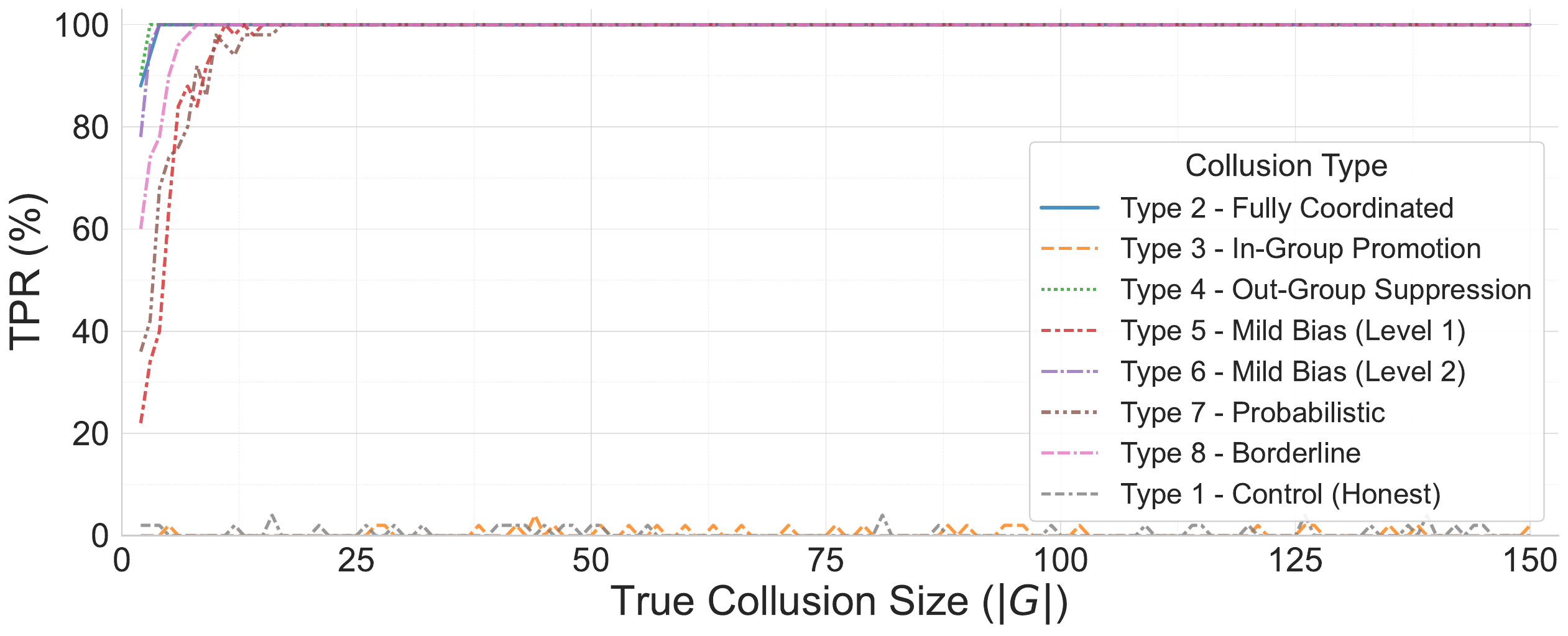}
    }\hfill
    \subfloat[]{
        \includegraphics[width=0.46\textwidth]{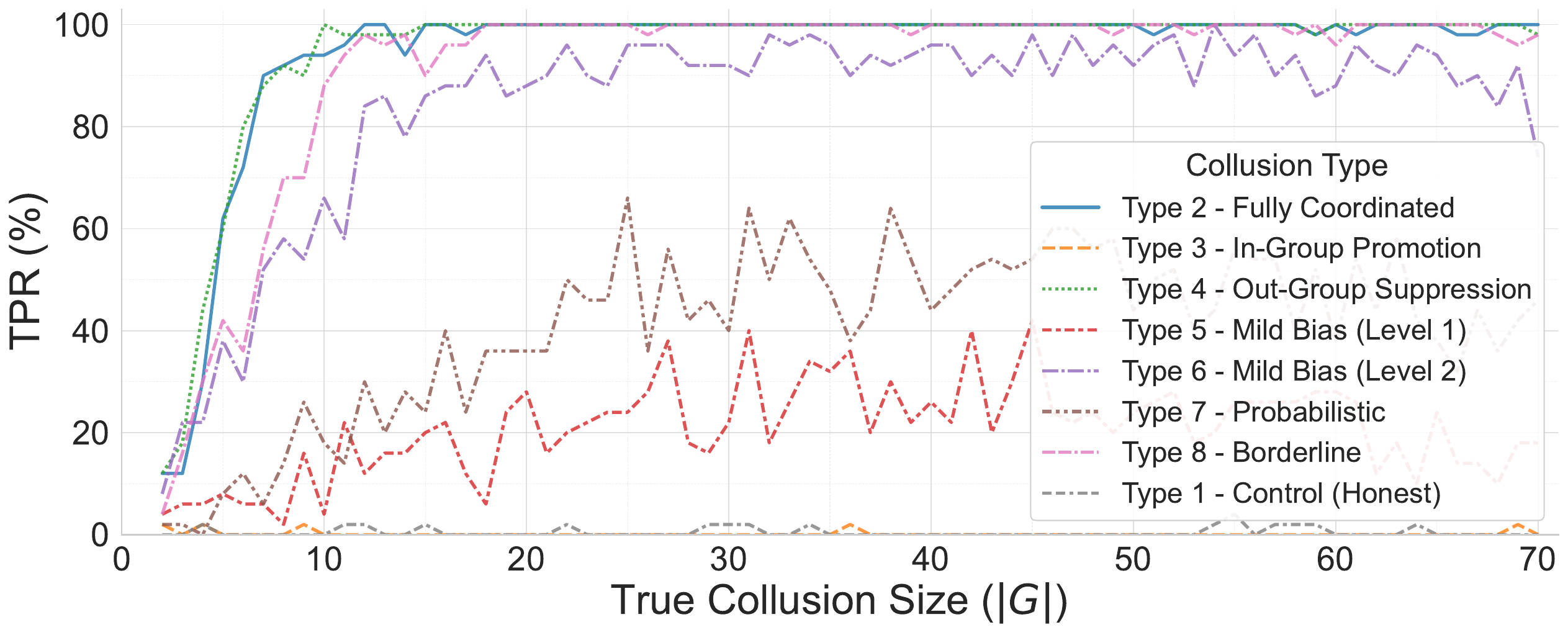}
    }
    
    \subfloat[]{
        \includegraphics[width=0.46\textwidth]{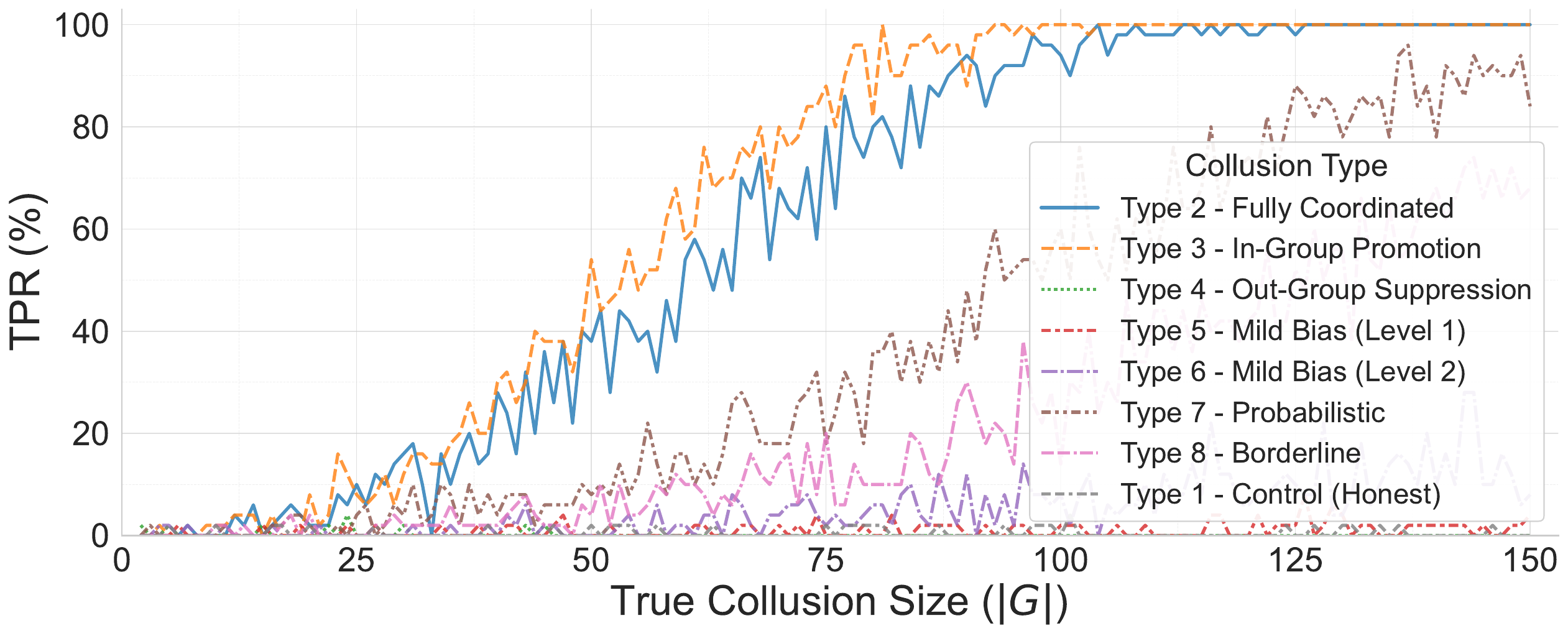}
    }\hfill
    \subfloat[]{
        \includegraphics[width=0.46\textwidth]{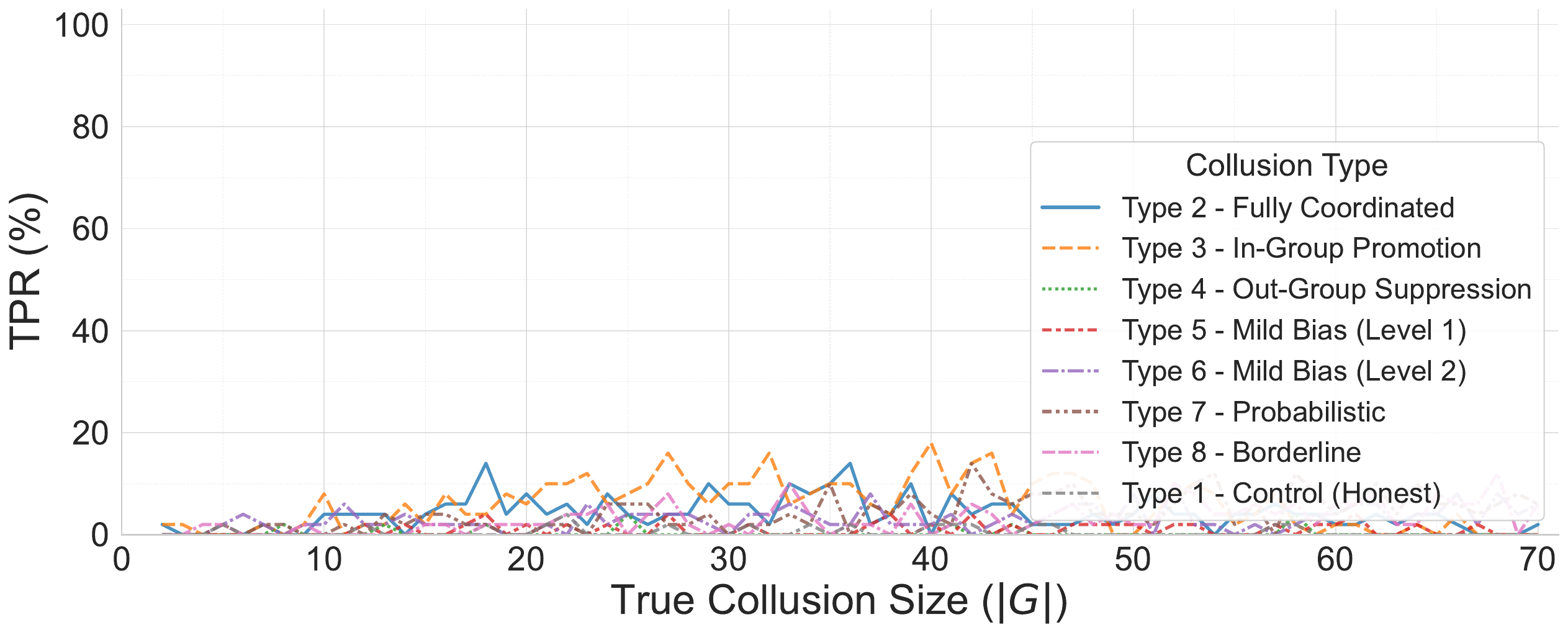}
    }

    \subfloat[]{
        \includegraphics[width=0.46\textwidth]{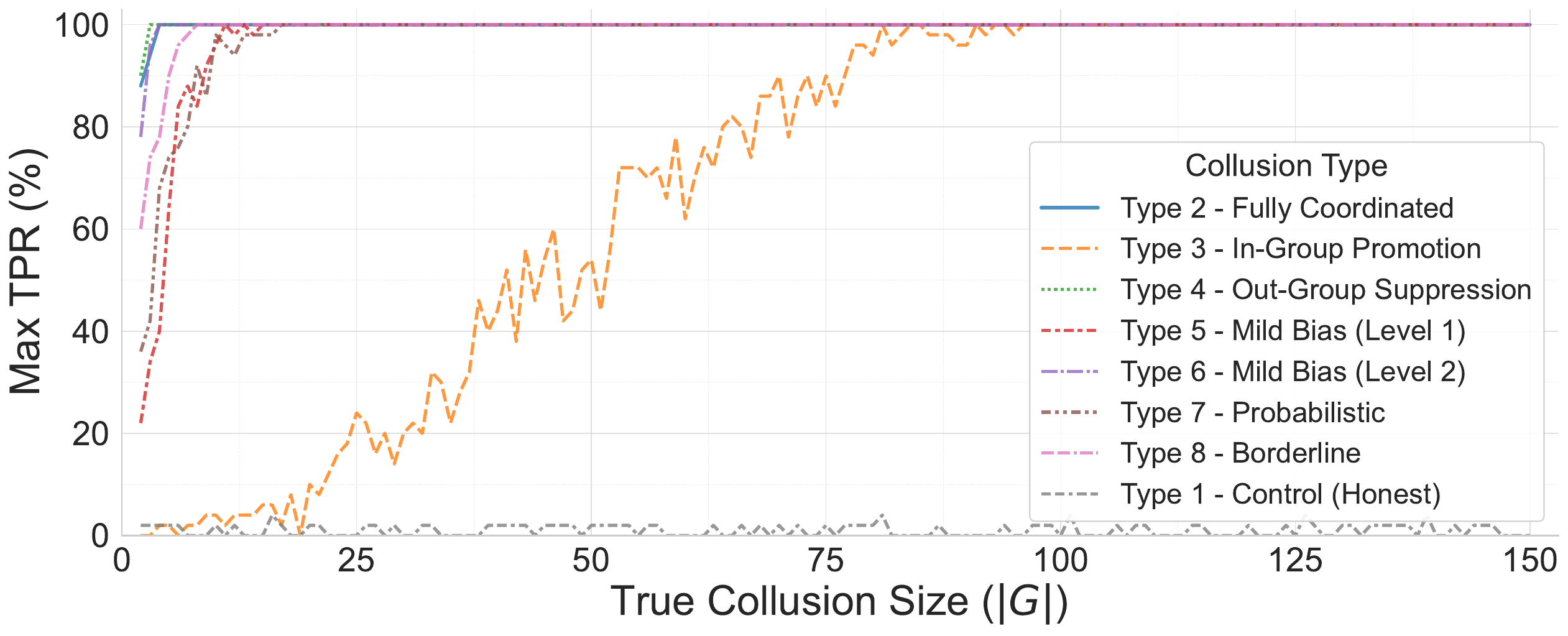}
    }\hfill
    \subfloat[]{
        \includegraphics[width=0.46\textwidth]{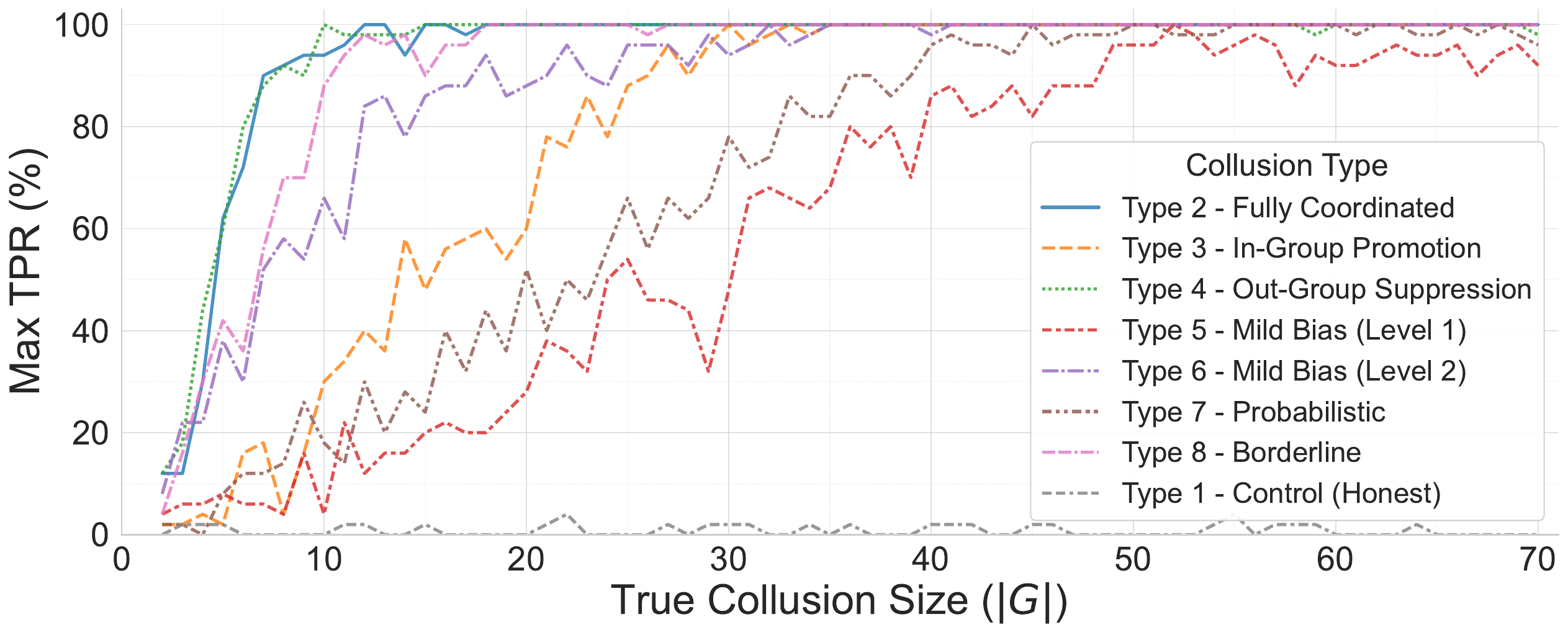}
    }
    
    \caption{True Positive Rate (TPR) analysis across datasets ICLR (left) vs. DPR (right). Rows (top to bottom): Algorithm 1, 2, 3 and combined. The legend at the top applies to all panels.}
    \label{fig:tpr_master_all}
\end{figure*}

In the continuous DPR setting (Figure~\ref{fig:tpr_master_all} b, d, f), the primary metrics demonstrate accelerated sensitivity. Algorithm~\ref{alg:algo2} (d) captures external suppression at minimal coalition sizes ($|G|\ge 10$) and maintains robust detection across most scenarios, with the notable exceptions of \textbf{Mild Coordinated Bias 1} and \textbf{Probabilistic Collusion}, which require slightly larger coalitions to be reliably flagged. Meanwhile, Algorithm~\ref{alg:algo1} (b) requires a greater critical mass to overcome variance, achieving strong detection rates for the majority of collusion types at $|G|\approx 40$. Nevertheless, it struggles to identify \textbf{Out-Group Suppression Only} and both variants of \textbf{Mild Coordinated Bias (1 and 2)}. In stark contrast, Algorithm~\ref{alg:algo3} (f) completely fails to demonstrate any detection capability in this continuous environment, remaining entirely unresponsive across all group sizes and collusion models.

\begin{figure*}[!ht]
    \centering
    \includegraphics[width=0.3\textwidth]{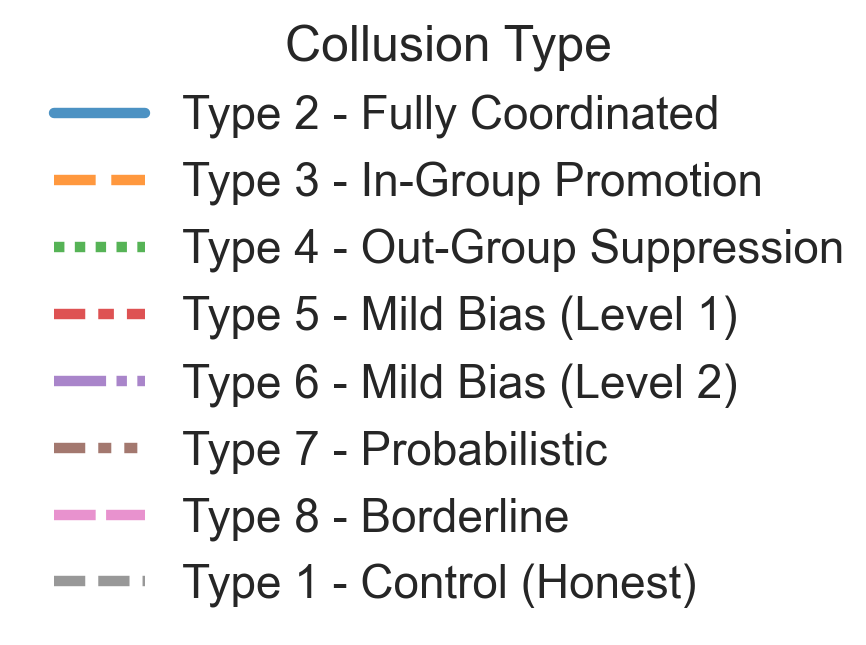} \\
    \subfloat[]{
        \includegraphics[width=0.46\textwidth]{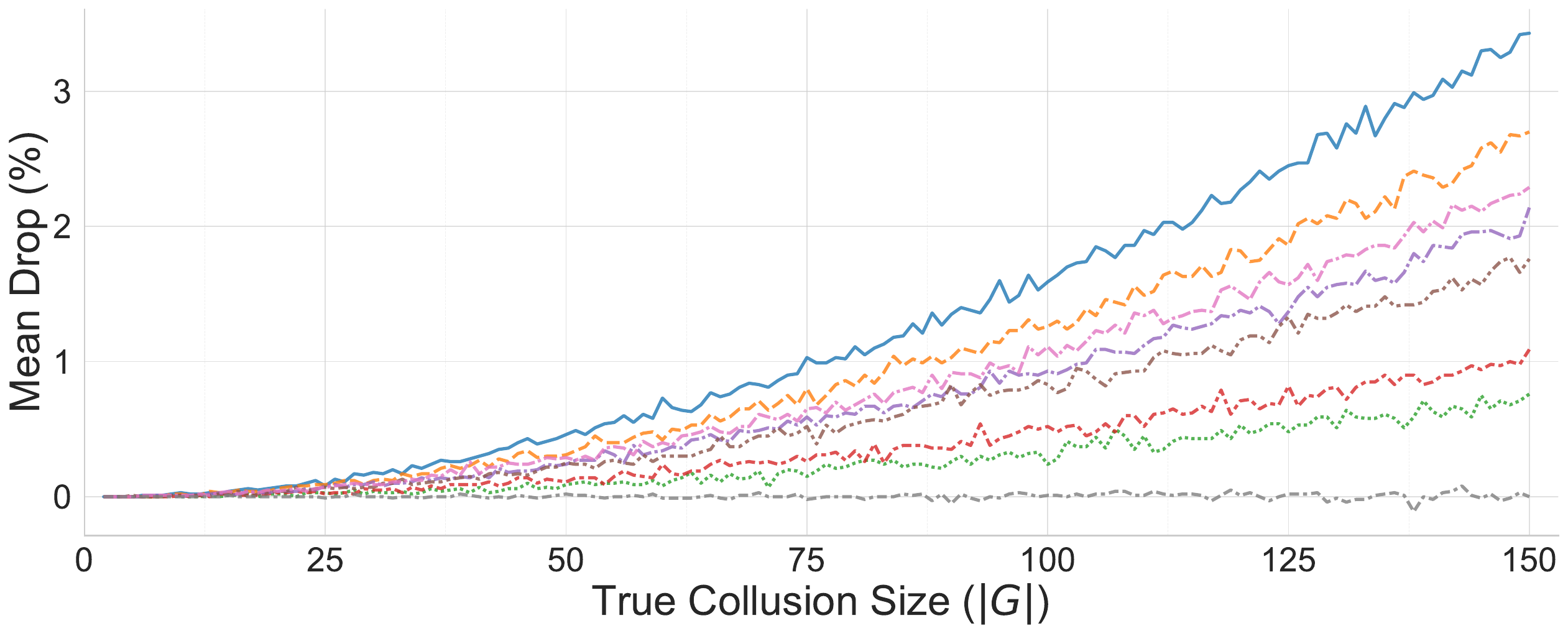}
    }\hfill
    \subfloat[]{
        \includegraphics[width=0.46\textwidth]{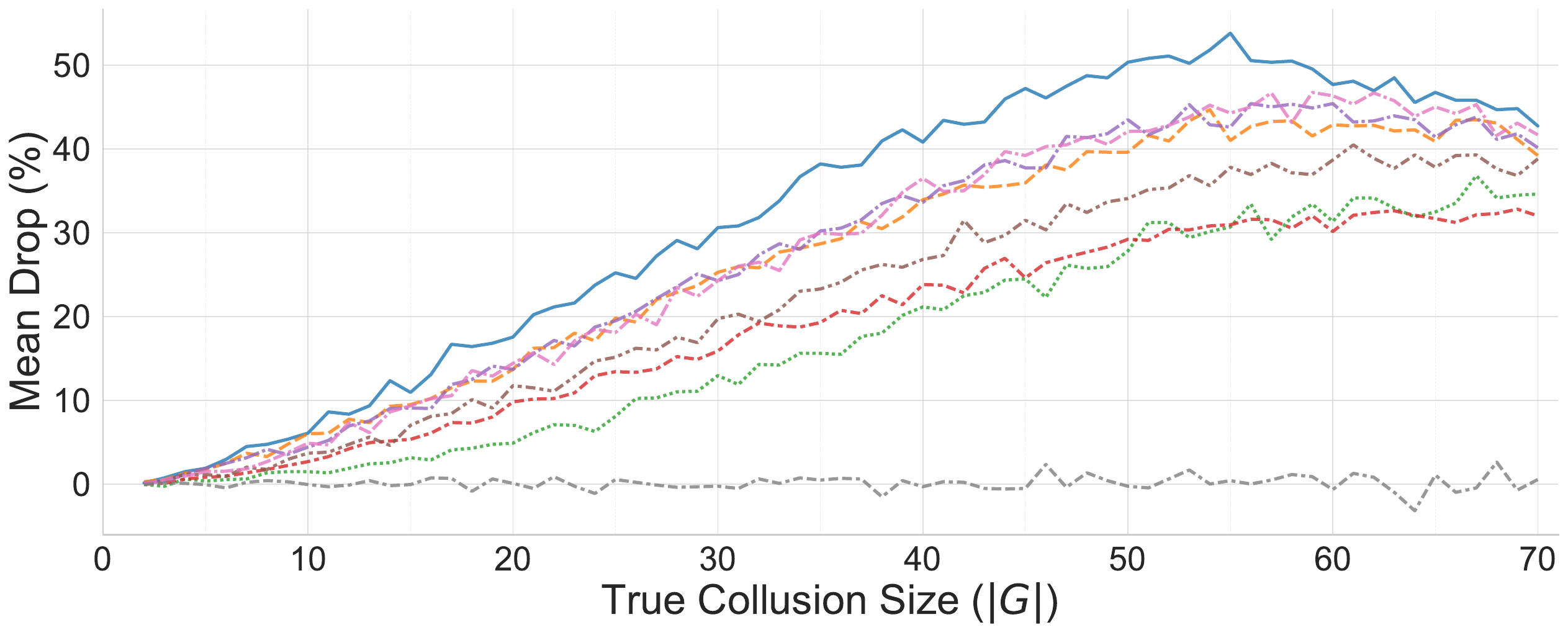}
    } \\
    \subfloat[]{
        \includegraphics[width=0.46\textwidth]{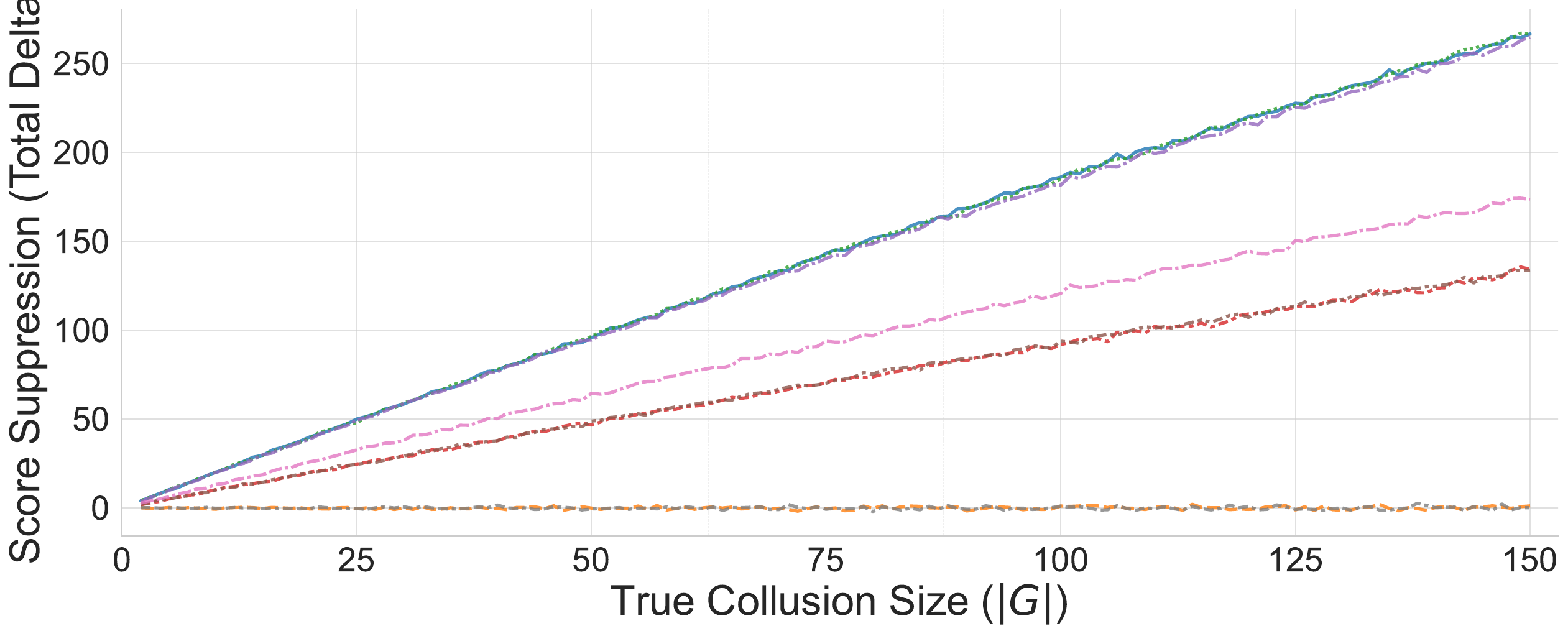}
    }\hfill
    \subfloat[]{
        \includegraphics[width=0.46\textwidth]{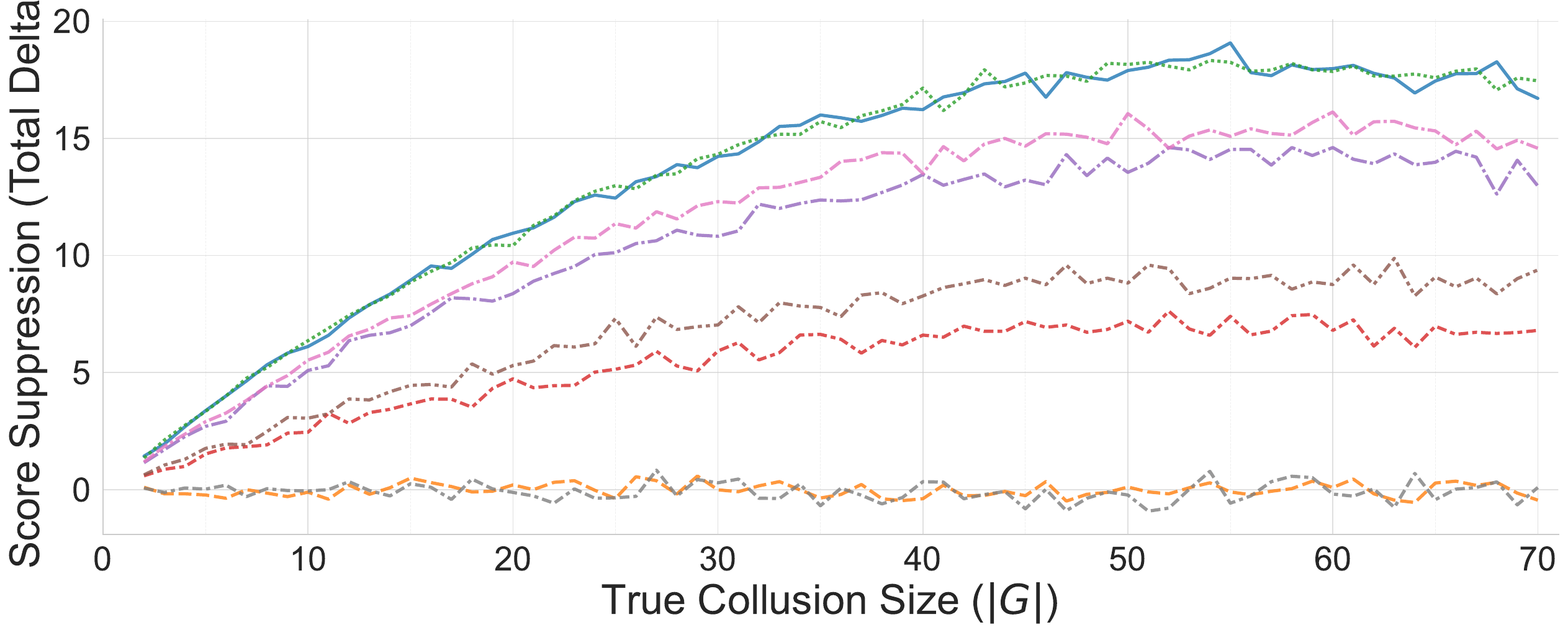}
    } \\
    \subfloat[]{
        \includegraphics[width=0.46\textwidth]{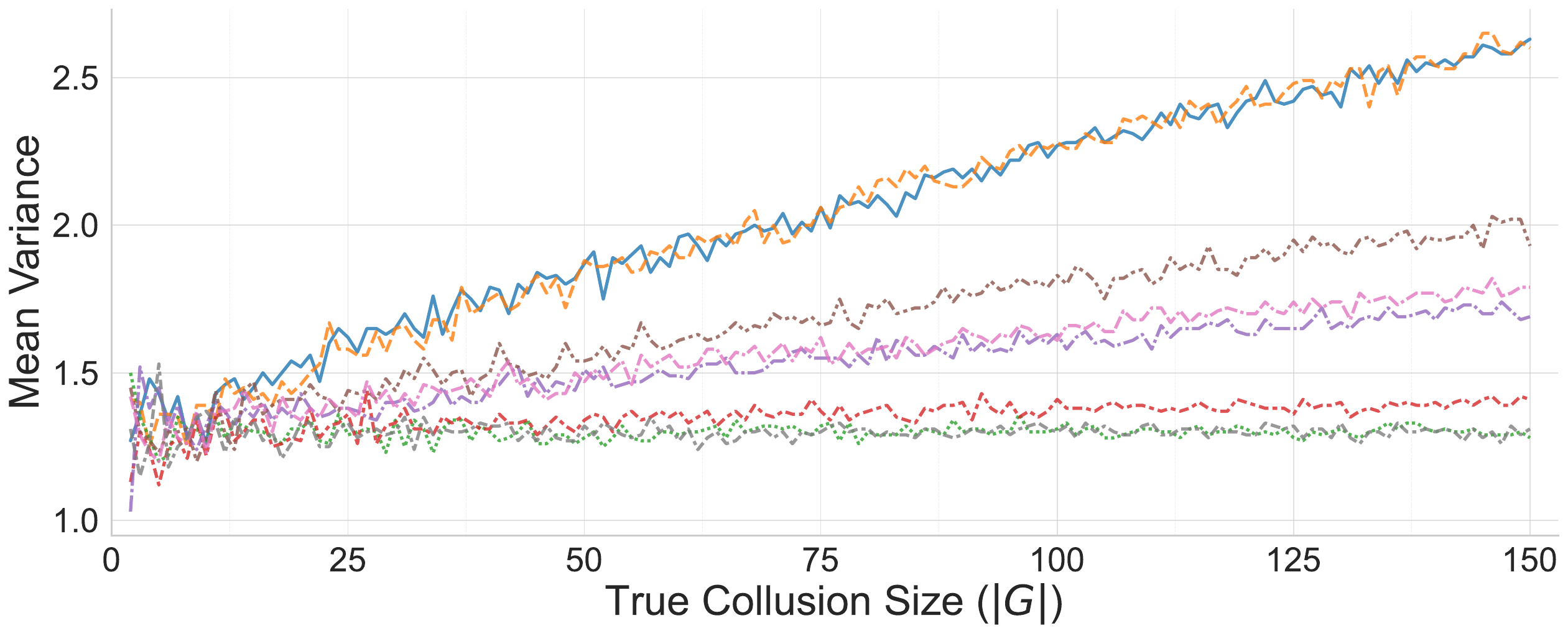}
    }\hfill
    \subfloat[]{
        \includegraphics[width=0.46\textwidth]{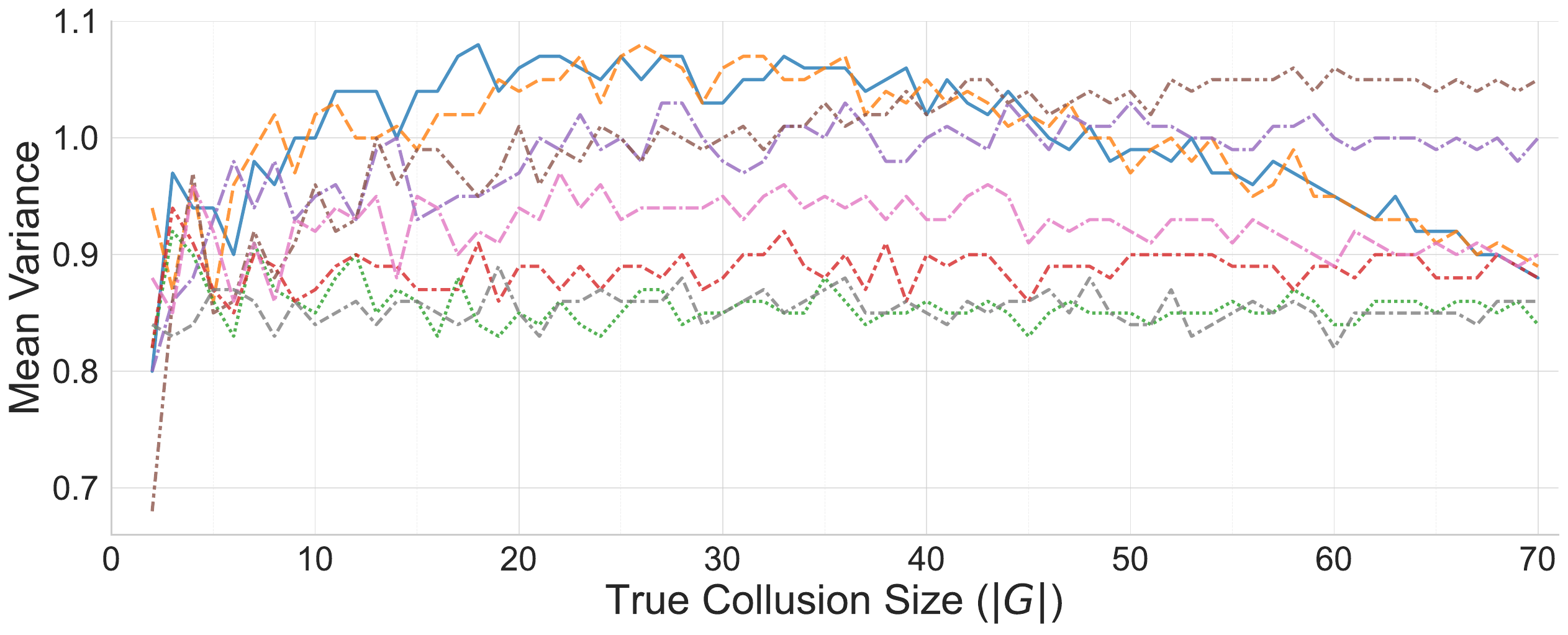}
    }
    \caption{Effect size dynamics per algorithm: ICLR (left) vs. DPR (right). Rows (top to bottom): Algorithms 1, 2, and 3. All plots share the legend at the top.}
    \label{fig:all_effect_sizes}
\end{figure*}

Collectively, as highlighted by the unified framework in Figure~\ref{fig:tpr_master_all} (h), integrating these metrics yields a compounded sensitivity that vastly outpaces any single standalone algorithm. The ensemble effectively shifts the reliable detection threshold leftward: by $|G|\approx 25$, the vast majority of collusion strategies are successfully and consistently identified. Even the most structurally subtle tactics, specifically \textbf{Mild Coordinated Bias 1} and \textbf{Probabilistic Collusion}, are forced above the detection threshold much earlier, proving the ensemble's capacity to amplify subtle, multi-dimensional manipulations into decisive systemic alerts.

\subsubsection{Algorithmic Sensitivities and Inherent Blind Spots:}
An analysis of the individual algorithms reveals distinct operational sensitivities and inherent blind spots, as corroborated by Figures~\ref{fig:all_effect_sizes} and~\ref{fig:tpr_master_all}. Notably, Algorithm~\ref{alg:algo1} successfully detects the \textbf{Out-Group Suppression Only} strategy. Although this tactic targets external papers, actively downgrading competitors inadvertently lowers the global acceptance threshold, thereby indirectly inflating the acceptance rate of the colluding group's own submissions. Conversely, Algorithm~\ref{alg:algo2} completely fails to identify \textbf{In-Group Promotion Only}, remaining unresponsive regardless of the colluding group's size. This is expected, as its detection mechanism is strictly formulated to measure anomalous downward shifts in external evaluations, rendering it blind to pure internal boosting.

While larger collusion rings are intuitively easier to detect, this holds only until they reach half the reviewer pool ($|G| \approx |\mathcal{R}| / 2$). Beyond this threshold, the coalition becomes the majority, dictating the baseline evaluation variance rather than deviating from it. This detection collapse mirrors fundamental limits in secure distributed computing, such as the BGW theorem~\cite{bgw1988}, which proves an honest majority ($t < n/2$) is strictly required to tolerate collusion. Fortunately, such massive rings are highly unrealistic in practice.

\subsection{False Positive Rate (FPR) Handle}
\label{subsec:soundness}

Although high detection sensitivity (TPR) is crucial, a robust detection framework must not misclassify innocent collaboration as malicious. To validate the structural soundness and discriminative power of our methodology, we rigorously evaluate its behavior under the null hypothesis (i.e., honest reviewer assignment).

\begin{figure}[!ht]
    \centering
    \includegraphics[width=0.95\linewidth]{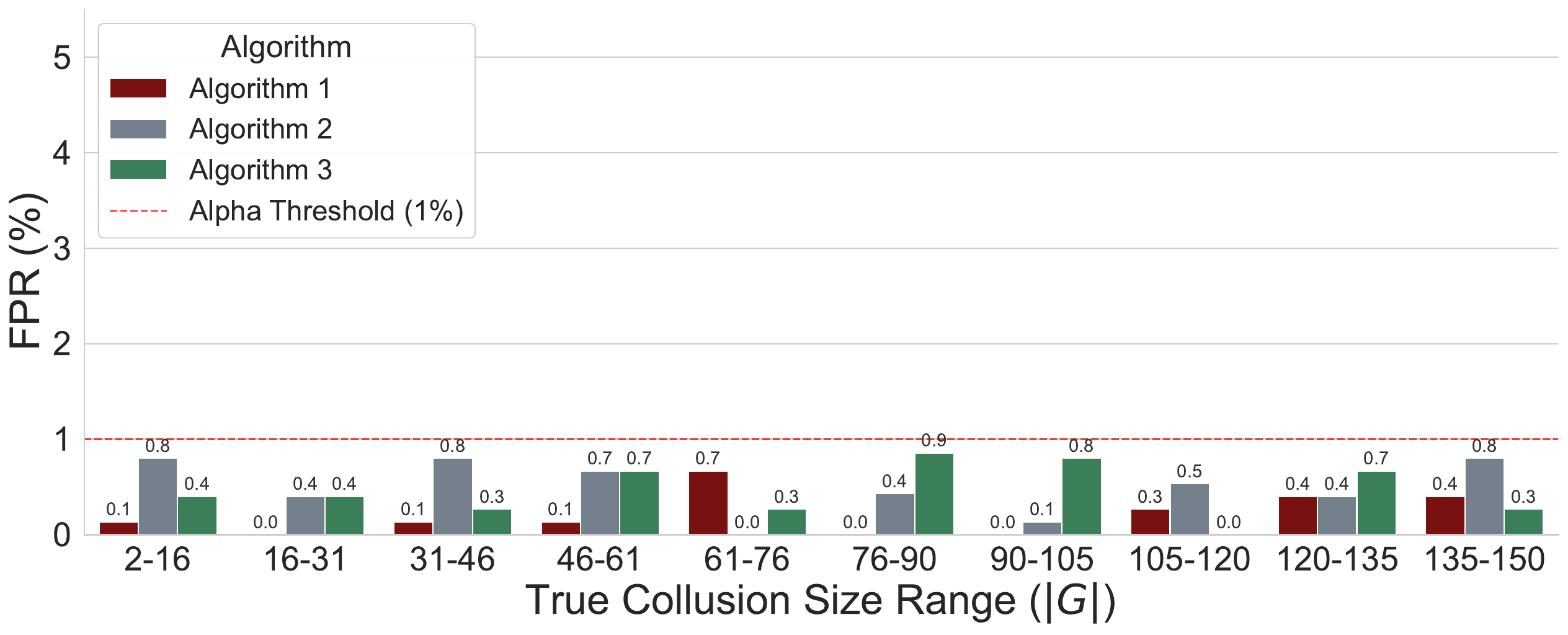}
    \caption{Empirical FPR Bounds for the ICLR dataset. The FPR for all algorithms remains strictly bounded beneath the significance threshold of $\gamma=0.01$.}
    \label{fig:fpr_bounds}
\end{figure}

As verified in Figure~\ref{fig:fpr_bounds}, the empirical False Positive Rate (FPR) across various group sizes remains strictly bounded beneath the committed significance threshold of $\gamma = 0.01$ for all algorithms. This confirms that our non-parametric permutation mechanics are structurally sound. The algorithms do not generate false alarms when processing natural venue noise or innocent reviewing patterns.

\subsection{Physical Constraints: Saturation Dynamics and Out-Group Depletion}
\label{subsec:saturation}

The compact nature of the DPR setting exposes unique physical and statistical constraints as the colluding group size expands. Specifically, we observe two distinct saturation phenomena that severely impact standalone detection metrics at extreme scales.

First, this saturation deeply affects the variance-based detection of Algorithm~\ref{alg:algo3}. In the extensive ICLR setting, large colluding coalitions generate sufficient statistical mass to produce massive, detectable variance spikes when their biased scores diverge from assigned honest reviewers. Conversely, in the smaller DPR pool, diminutive rings rapidly saturate the review assignments of their own papers. This high in-group density effectively eliminates honest external baselines on target submissions, dropping internal scoring variance to near zero and rendering Algorithm~\ref{alg:algo3} unresponsive.

Second, we highlight the \textit{Out-Group Depletion} phenomenon visible in the effect size analysis of Figure~\ref{fig:all_effect_sizes}(d), which directly limits the external suppression measured by Algorithm~\ref{alg:algo2}. As a colluding ring scales to encapsulate a significant majority of the venue ($|G|> 55$), the total external target pool available for sabotage shrinks proportionally. Consequently, the cumulative suppression impact rounds off into a parabolic saturation curve and begins to decline. 

Ultimately, these dual dynamics demonstrate that even aggressive systemic collusion is inherently bounded by finite venue parameters and individual reviewing workload limits. Furthermore, these environmental vulnerabilities strictly justify our multi-metric ensemble design, ensuring the combined suite (Figure~\ref{fig:tpr_master_all}) compensates optimally where individual algorithms saturate or fail.

\FloatBarrier

\section{The Embedding Based Discovery Framework}
\label{sec:discovery_framework}

\subsection{Zero-Knowledge Detection and the Combinatorial Explosion}

In real-world peer review, conference organizers possess the review evaluation matrix $E$ but lack prior knowledge of which reviewers might be colluding. Therefore, identifying a covert collusion ring by exhaustively evaluating all reviewer subsets is computationally intractable ($\mathcal{O}(2^{|\mathcal{R}|})$), necessitating a targeted heuristic approach to identify viable candidate groups.

However, organized collusion is rarely spontaneous. To successfully manipulate outcomes without raising immediate suspicion, colluding actors must inherently share a foundational common denominator, such as overlapping research interests or institutional affiliations, to successfully coordinate.

Leveraging this insight, we introduce a coarse-to-fine discovery pipeline. First, we partition the reviewer pool into non-overlapping candidate clusters by capturing their semantic and institutional connections within a dense embedding space. Each cluster is then independently evaluated using our diagnostic algorithms to quantify its deviation from expected scoring behaviors. Once the most suspicious initial clusters are isolated, we apply an iterative, greedy refinement process to discard innocent bystanders and extract the true colluding core. Finally, to ensure robustness and mitigate algorithmic biases, the distinct outputs from each objective function are unified, culminating in a highly precise and statistically rigorous set of identified colluders.

\subsection{Step 1: Semantic and Geometric Partitioning}

To capture genuine academic proximity and overcome the ease of hiding explicit co-authorship ties, we rely on intrinsic reviewer metadata. We extract the textual metadata of all reviewers, specifically their research expertise keywords and historical institutional domains. To prevent spurious correlations, we rigorously preprocess this metadata. Notably, generic email domains (e.g., \textit{gmail.com}) are replaced with randomized sequences. This ensures the clustering algorithm does not artificially group reviewers simply due to the widespread use of commercial email providers or the absence of institutional addresses, thereby forcing the embeddings to strictly reflect genuine academic alignments.

\begin{figure}[!ht]
    \centering
    \subfloat[Semantic Clustering Topology]{
        \includegraphics[width=0.46\textwidth]{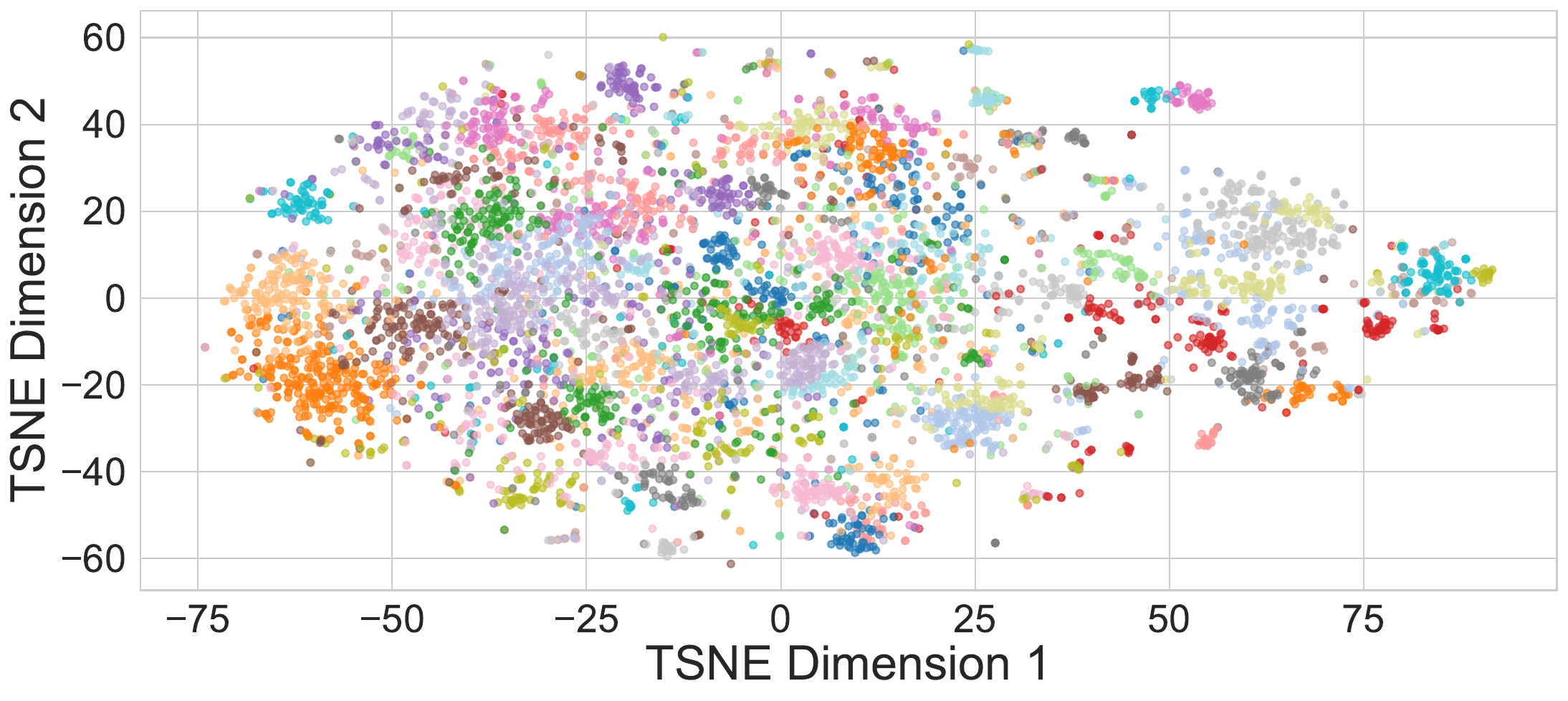}
    }
    \hfill
    \subfloat[Elbow and Silhouette Score\protect\footnotemark]{
        \includegraphics[width=0.46\textwidth]{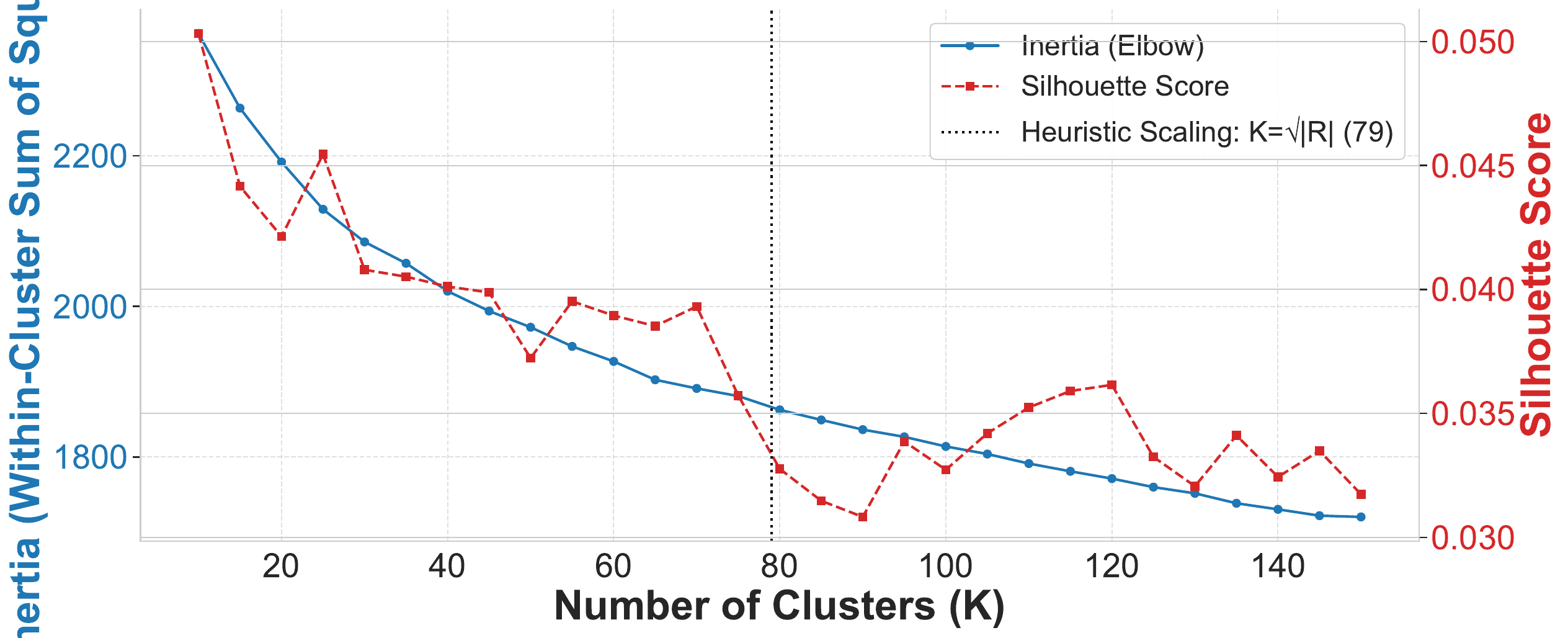}
    }
    
    \caption[Embedding Based Framework Initialization]{Embedding Based Framework Initialization: (a) A two-dimensional projection of the semantic reviewer embeddings ($K=79$). (b) The empirical justification for $K$ selection in ICLR via the Elbow method and Silhouette scores.}
    \label{fig:discovery_initialization}
\end{figure}

These features are converted into dense vector embeddings using a robust, pre-trained transformer model (\textit{all-mpnet-base-v2})~\cite{reimers2019}, applying $L_2$ normalization to ensure Euclidean distance properly reflects cosine similarity. We project them into a unified semantic space (as shown in Figure~\ref{fig:discovery_initialization}(a)). 

Traditional community detection often relies on explicit co-authorship graphs. However, these networks are inherently limited; they only capture formally declared collaborations and can be easily bypassed by colluders who coordinate through informal or hidden social channels (e.g., unpublished institutional ties or "friends-of-friends" networks). Unlike formal co-authorship networks, this semantic baseline prevents malicious actors from easily obfuscating their relationships. While malicious actors can easily obscure or omit declaring past co-authorships to hide a relationship, they cannot effectively participate in a targeted reviewing ring without sharing a common, documentable baseline of academic expertise.

To transition from these continuous semantic embeddings into discrete candidate groups, we apply K-Means clustering\footnote{K-Means partitions the embedding space by grouping points around their nearest centroid, effectively isolating semantically cohesive communities based on geometric proximity.} to the unified space~\cite{macqueen1967}. A fundamental challenge in this unsupervised partitioning is determining the appropriate granularity. To ensure our framework is adaptable to any conference, the number of clusters, $K$, is dynamically scaled relative to the reviewer pool size according to the widely-adopted heuristic rule $K = \lfloor\sqrt{|\mathcal{R}|}\rfloor$. This sub-linear scaling elegantly balances a fundamental structural trade-off: it prevents the over-fragmentation of natural research domains while maintaining clusters small enough to be computationally tractable for our localized optimization algorithms.

In the context of the ICLR dataset, which comprises $|\mathcal{R}| = 6385$ unique reviewers, our scaling heuristic prescribes $K = 79$. To verify that this mathematical rule-of-thumb genuinely reflects the underlying semantic topology of the conference, we turn to our empirical metrics. As depicted in Figure~\ref{fig:discovery_initialization} (b), the Inertia curve exhibits a distinct diminishing rate of decrease exactly as we approach $K \approx 79$, marking the transition where additional clusters provide only marginal gains. Simultaneously, the Silhouette score reaches a local maximum and stabilizes at this precise granularity. This striking alignment between the theoretical scaling rule and the empirical data firmly justifies $K = 79$ as the optimal partition point. Ultimately, this ensures the resulting base clusters strike the perfect balance: they are sufficiently dense to capture highly correlated research communities, yet broad enough to avoid artificially fracturing natural academic domains into fragmented, non-informative noise.

\footnotetext{The \textbf{Elbow method} evaluates the within-cluster variance (inertia), identifying the point of diminishing returns (the "elbow") when increasing $K$. The \textbf{Silhouette score} measures cluster quality by comparing intra-cluster cohesion to inter-cluster separation, where values closer to 1 indicate dense and well-separated clusters.}

\subsection{Step 2: Decoupled Candidate Prioritization}
\label{subsec:candidate_prioritization}

Instead of evaluating the candidate clusters using a single unified metric, our framework processes each diagnostic dimension independently. For a given candidate group $G$, we define three tailored objective functions $\mathcal{Q}_1, \mathcal{Q}_2, \mathcal{Q}_3$ corresponding to our core diagnostic algorithms\footnote{Let $\mathcal{A}_a(G)$ denote the scalar outputted by algorithm $a$ for a given group $G$.}. These objective functions normalize the raw outputs into comparable scales while applying distinct structural modifiers and a shared size-based penalty (the tax rate parameter, $\lambda \cdot |G|$) to strictly prevent the absorption of innocent reviewers:

\begin{enumerate}
    \item \textbf{In-Group Promotion Objective ($\mathcal{Q}_1$):} Since $\mathcal{A}_1(G)$ naturally outputs a fractional drop in acceptance rates (e.g., $0.05$ for a $5\%$ drop), we multiply the algorithm result by $100$. This straightforward scaling maps the fractional rate to a magnitude where the standard linear tax penalty ($\lambda$) can effectively regularize the score:
    \begin{equation}
        \mathcal{Q}_1(G) = 100 \cdot \mathcal{A}_1(G) - \lambda \cdot |G|
    \end{equation}
    
    \item \textbf{Out-Group Suppression Objective ($\mathcal{Q}_2$):} The raw cumulative suppression sum is squared to exponentially reward dense, coordinated impact. However, to prevent larger groups from naturally inflating this sum simply due to their size, we normalize it by $|G|^{1.1}$ (a slightly super-linear scaling) before applying the standard final tax:
    \begin{equation}
        \mathcal{Q}_2(G) = \frac{(\mathcal{A}_2(G))^2}{|G|^{1.1}} - \lambda \cdot |G|
    \end{equation}
    
    \item \textbf{Scoring Variance Objective ($\mathcal{Q}_3$):} Maintaining high internal scoring variance becomes statistically harder as a group grows. Therefore, we scale the raw variance by $1.5 \cdot \sqrt{|G|}$ to actively reward larger groups that still manage to exhibit strong anomalous variance. Additionally, because a variance signal inherently requires a critical mass of members to become meaningful, we apply a more forgiving size penalty here, effectively reducing the tax to one-third ($-\frac{\lambda}{3} \cdot |G|$):
    \begin{equation}
        \mathcal{Q}_3(G) = 1.5 \cdot \sqrt{|G|} \cdot \mathcal{A}_3(G)  - \frac{\lambda}{3} \cdot |G|
    \end{equation}
\end{enumerate}

The prioritization engine evaluates these three functions across all generated base clusters $\mathcal{C}$. For each algorithm $a \in \{1, 2, 3\}$, the clusters are sorted independently based on their respective $\mathcal{Q}_a(C_i)$ scores. The top-$b$ most suspicious clusters for each algorithm (where $b$ is the configurable search breadth parameter) are then selected to undergo individual greedy refinement.

\subsection{Step 3: Decoupled Iterative Refinement and Consensus Formations}
\label{subsec:iterative_refinement}

Following the prioritization phase, the framework refines the top-$b$ candidate clusters independently for each algorithm $a \in \{1, 2, 3\}$. Each group is optimized locally to maximize its corresponding objective function $\mathcal{Q}_a$.

\textbf{Candidate Pool:} Because a collusion ring may cross strict geometric boundaries, we expand the search space. The candidate pool for each cluster includes its current members plus all reviewers residing in the top $\left\lceil 20\% \right\rceil$ nearest adjacent clusters (measured by Euclidean distance between cluster centroids) within the semantic space.

\textbf{The $+1/-1$ Search Dynamics:} In each iteration, the engine evaluates two potential modifications:
\begin{enumerate}
    \item \textbf{Pruning ($-1$):} Removing a current member from the group.
    \item \textbf{Absorption ($+1$):} Adding a reviewer from the candidate pool.
\end{enumerate}

Similar to a Hill Climbing search, the algorithm acts greedily, executing the single operation that yields the maximum improvement to $\mathcal{Q}_a(G)$, provided the gain exceeds a minimal threshold $\epsilon$:
\begin{equation}
    \Delta \mathcal{Q}_a > \epsilon
\end{equation}

\textbf{Tabu State Memory:} To avoid infinite cyclic loops, group configurations are hashed after every step; actions leading to previously visited states are rejected.

After independently refining all $b$ candidate clusters for each algorithm, the process yields one absolute best-scoring group per algorithm: $G_1$, $G_2$, and $G_3$. Rather than relying on a single metric, we combine these groups into three consensus formations to balance precision and recall:

\begin{itemize}
    \item \textbf{Intersection Group ($G_{\text{inter}}$):} Defined as $G_1 \cap G_2 \cap G_3$. This isolates the dense core agreed upon by all algorithms, maximizing precision.
    \item \textbf{Majority Vote Group ($G_{\text{maj}}$):} Comprising reviewers present in at least two groups, calculated as $(G_1 \cap G_2) \cup (G_1 \cap G_3) \cup (G_2 \cap G_3)$. This offers a balanced diagnostic compromise.
    \item \textbf{Union Group ($G_{\text{union}}$):} The complete union $G_1 \cup G_2 \cup G_3$. This maximizes recall to capture peripheral colluders, albeit at a higher false-alarm rate.
\end{itemize}

The complete pseudo-code detailing this decoupled discovery pipeline is provided in Algorithm~\ref{proc:discovery_new}.

\subsection{Step 4: Multi-Algorithmic Statistical Validation}
\label{subsec:statistical_validation}

Searching numerous potential sub-graphs introduces the Multiple Comparisons Problem, inflating the probability of false positives under standard nominal thresholds. To prevent false accusations, we strictly control the Family-Wise Error Rate (FWER). The FWER is defined as the probability of making at least one Type I error (i.e., a false positive) across the entire 'family' of evaluated groups. By bounding the FWER, we provide a mathematical guarantee that the probability of falsely flagging even one honest group remains below $\gamma$.

As the consensus groups are newly formed amalgams of the individual searches, we must re-evaluate their significance across all metrics. To guarantee FWER control, we apply a permutation test that compares these new groups against the maximum scores expected under random assignment. For each evaluated consensus group $G \in \{G_{\text{inter}}, G_{\text{maj}}, G_{\text{union}}\}$, we compute its diagnostic scores $\mathcal{A}_a(G)$ for $a \in \{1, 2, 3\}$. We execute $T$ permutation iterations. In each iteration $t$, the engine generates $K$ random reviewer groups $R_j^{(t)}$ of size $|G|$, sampled uniformly without replacement from the entire pool\footnote{Generating $K$ groups mirrors the multiple-comparison burden of evaluating the initial semantic clusters, ensuring the extreme-value null distribution accurately reflects the search space scope.}. To simulate the worst-case null noise, we record the maximum score per metric:
\begin{equation}
    M_{a}^{(t)} = \max_{j \in \{1..K\}} \mathcal{A}_a(R_j^{(t)}) \quad \text{for } a \in \{1, 2, 3\}
\end{equation}

This constructs three independent extreme-value null distributions. The FWER-adjusted $p$-values for the specific group $G$ are then calculated by mapping observed scores directly against these thresholds:
\begin{equation}
    p_a(G) = \frac{1 + \sum_{t=1}^T \mathbb{I}\left[M_{a}^{(t)} \ge \mathcal{A}_a(G)\right]}{T+1} \quad \text{for } a \in \{1, 2, 3\}
\end{equation}

To neutralize single-metric anomalies, the framework enforces a majority-voting rule. Group $G$ is declared FWER-significant ($\text{Is\_Sig}_G$) at level $\gamma$ if it triggers at least two diagnostic dimensions:
\begin{equation}
    \text{Is\_Sig}_G = \left( \sum_{a=1}^3 \mathbb{I}[p_a(G) < \gamma] \ge 2 \right)
\end{equation}

Ultimately, achieving a high frequency of $\text{Is\_Sig}_G = \text{True}$ is the crucial measure of algorithmic validity, demonstrating the system reliably extracts malicious coalitions while proving such extreme statistical signatures are highly improbable under honest assignment.

\begin{algorithm}[!ht]
\begin{small}
    \caption{Decoupled Embedding Based Discovery Pipeline}
    \label{proc:discovery_new}
    \textbf{Input:} Reviewer metadata $X$, Evaluation matrix $E$, Tax rate $\lambda$, Minimum improvement $\epsilon$, Search breadth $b$, Trials $T$, Significance level $\gamma$.
    
    \textbf{Output:} The three consensus groups and their significance flags ($\text{IsSig}_G$).
    \begin{algorithmic}[1]
        \State Embed $X$ using \textit{all-mpnet-base-v2} and mask generic emails.
        \State $\mathcal{C} \gets \textsc{K-Means}(X, K)$ where $K = \lfloor\sqrt{|\mathcal{R}|}\rfloor$
        
        \For{\textbf{each} algorithm $a \in \{1, 2, 3\}$}
            \State $\mathcal{C}_{top}^{(a)} \gets \text{Top-$b$ clusters from } \mathcal{C} \text{ based on } \mathcal{Q}_a$
            \State $G_a \gets \emptyset, \quad Q_{max} \gets -\infty$
            
            \For{\textbf{each} $C_i \in \mathcal{C}_{top}^{(a)}$}
                \State $G_{curr} \gets C_i, \quad \textit{TabuList} \gets \{ \textsc{Hash}(C_i) \}, \quad \textit{Improved} \gets \textbf{True}$
                \State $\mathcal{F} \gets \textsc{NearestClustersPool}(G_{curr}, 20\%) \cup G_{curr}$
                
                \While{\textit{Improved} \textbf{and} $|G_{curr}| \ge 3$}
                    \State $G_{test} \gets \text{Valid } (+1/-1) \text{ modification maximizing } \mathcal{Q}_a$
                    \Statex \hfill $\text{where } \textsc{Hash}(G_{test}) \notin \textit{TabuList}$

                    \If{$\mathcal{Q}_a(G_{test}) - \mathcal{Q}_a(G_{curr}) > \epsilon$}
                        \State $G_{curr} \gets G_{test}, \quad \textit{TabuList}.\text{add}(\textsc{Hash}(G_{curr}))$
                    \Else
                        \State $\textit{Improved} \gets \textbf{False}$
                    \EndIf
                \EndWhile
                
                \If{$\mathcal{Q}_a(G_{curr}) > Q_{max}$}
                    \State $Q_{max} \gets \mathcal{Q}_a(G_{curr}), \quad G_a \gets G_{curr}$
                \EndIf
            \EndFor
        \EndFor
        
        \State $G_{\text{inter}} \gets G_1 \cap G_2 \cap G_3$
        \State $G_{\text{maj}} \gets (G_1 \cap G_2) \cup (G_1 \cap G_3) \cup (G_2 \cap G_3)$
        \State $G_{\text{union}} \gets G_1 \cup G_2 \cup G_3$
        
        \For{\textbf{each} $G \in \{G_{\text{inter}}, G_{\text{maj}}, G_{\text{union}}\}$}
            \For{$t = 1$ to $T$}
                \State Generate $K$ random groups $\{R_1^{(t)}, \dots, R_K^{(t)}\}$ of size $|G|$
                \State $M_a^{(t)} \gets \max_{j \in \{1..K\}} \mathcal{A}_a(R_j^{(t)}) \quad \textbf{for all } a \in \{1, 2, 3\}$
            \EndFor
            
            \State $p_a(G) \gets \frac{1 + \sum_{t=1}^T \mathbb{I}[M_a^{(t)} \ge \mathcal{A}_a(G)]}{T+1} \quad \textbf{for all } a \in \{1, 2, 3\}$
            \State $\textit{IsSig}_G \gets \sum_{a=1}^3 \mathbb{I}[p_a(G) < \gamma] \ge 2$
        \EndFor
        
        \State \Return $\{ (G_{\text{inter}}, \textit{IsSig}_{\text{inter}}), (G_{\text{maj}}, \textit{IsSig}_{\text{maj}}), (G_{\text{union}}, \textit{IsSig}_{\text{union}}) \}$
    \end{algorithmic}
\end{small}
\end{algorithm}

\FloatBarrier

\section{Discovery Evaluation in Semi-Synthetic Environments}
\label{sec:evaluation}

Having established the Decoupled Embedding Based Discovery Framework, we must empirically evaluate its classification performance, statistical power, and computational scalability. To achieve this, we benchmark our approach using the ICLR dataset (which previously served as the foundation for our controlled simulations in Section~\ref{sec:simulation_results}). Beyond the standard bipartite graph of submissions and review scores, this dataset provides us rich semantic metadata regarding the authors, encompassing their specific research domains, institutional affiliations, and professional histories. Our discovery evaluation focuses exclusively on the ICLR dataset to accurately reflect the semantic complexity of large-scale venues. The clustering framework relies on rich, unstructured author metadata to extract latent relationships. The DPR dataset, being a controlled and localized experiment, simply lacks the semantic density and institutional diversity required to rigorously stress-test this pipeline.

However, we encounter the same operational constraint highlighted in Section~\ref{sec:simulation_results}: the exact identities of the reviewers assigned to specific papers are classified, restricted exclusively to the conference program chairs. To overcome this limitation for our experimental validation, we randomly assign the anonymized reviewer IDs to known author profiles during the collusion injection phase. This approximation relies on the standard and widely accepted assumption that in top-tier academic venues, the pool of submitting authors and the pool of active peer reviewers are practically identical, allowing us to evaluate the framework's clustering and discovery capabilities on a highly realistic semantic landscape. To ensure the reproducibility of our results, the source code used for all experiments will be made publicly available.

\subsection{Experimental Setup: The Hide-and-Seek Protocol and Computational Feasibility}
\label{subsec:experimental_setup}

Evaluating collusion detection in real-world datasets suffers from ground-truth scarcity: conference organizers lack absolute labels of covert malicious behavior. To calculate rigorous Precision and Recall metrics, we constructed a controlled ``Hide-and-Seek'' simulation laboratory within the ICLR environment. 

\textbf{Hide (Adversarial Injection):} To mimic real-world data imperfections, we inject minor semantic noise. The adversary injects zero-mean Gaussian noise ($\sigma=0.0001$, calibrated to prevent identical embedding vectors without destroying the underlying semantic geometry) into the reviewers' semantic embeddings, creating an obfuscated clustering space. To form a logically coherent yet hidden collusion ring of target size $\ell$, the adversary strategically selects a random base cluster, expands it by absorbing geometric neighbors until the pool exceeds $\ell$, and randomly samples exactly $\ell$ reviewers. Finally, the \texttt{FULLY\_COORDINATED\_COLLUSION} manipulation is deployed for this specific group into the review matrix.

\textbf{Seek (Algorithmic Recovery):} Completely blinded to the ground-truth memberships and semantic perturbations, the discovery engine processes the corrupted matrix. Its goal is to blindly reconstruct the hidden ring using our multi-algorithmic consensus formations.

To rigorously validate the pipeline, we swept across injected ring sizes $\ell \in [2, 150]$. To eliminate stochastic variance caused by noise and K-Means initialization, we executed 20 independent full-scale iterations per target size, generating thousands of unique diagnostic trials.

Subjecting our localized greedy search to thousands of iterations naturally highlights a computational bottleneck. To make this massive combinatorial sweep computationally feasible, we engineered strict technical optimizations, including semantic pre-computation and graph-based marginal caching:
\begin{enumerate}
    \item \textbf{Parallel Universe Pre-computation:} Instead of dynamically performing K-Means for every experiment, we pre-computed the semantic clusters and centroids for all 20 iterations upfront. The $i$-th evaluation strictly reuses the $i$-th pre-computed geometry, guaranteeing controlled semantic divisions while eliminating redundant overhead.
    \item \textbf{Graph-Based Marginal Caching:} Evaluating thousands of overlapping states differing by a single $+1/-1$ action using full pandas matrix operations is strictly intractable. We engineered fast dictionary-based scoring variants that construct native dependency pointers between reviewers, submissions, and scores. This allows the framework to instantaneously compute localized marginal deltas rather than rescanning the entire evaluation matrix.
\end{enumerate}

With these optimizations, we executed the full diagnostic pipeline on the ICLR dataset ($|\mathcal{R}| = 6385$). To balance exhaustive anomaly detection with computational feasibility, the framework was deployed with empirically calibrated hyperparameters ($b=10$, $\lambda=0.4$, $\epsilon=0.05$, $T=200$, $\gamma=0.01$). Specifically, we set a search breadth of $b=10$ (capturing the top $\sim 12\%$ of the $K=79$ base clusters, from which the globally optimal group is highly likely to emerge), a penalty tax of $\lambda=0.4$ (calibrated to be small relative to the objective gain of absorbing a true colluder, yet significant enough to remain a relevant deterrent against bloated groups), a minimum gain threshold of $\epsilon=0.05$ (halting the search engine when improvements become minor and are likely mere structural noise), and $T=200$ permutation trials (a sample size substantial enough to validate the rigorous statistical tests without excessively burdening the overall runtime) at a stringent significance level of $\gamma=0.01$ (enforcing a very high certainty threshold, given the dramatic implications and reputational risks of falsely accusing researchers of collusion). Under this configuration, the complete diagnostic sweep requires approximately a day and a half on standard hardware. This runtime remains highly practical for post-conference auditing, where organizers typically possess weeks to verify review integrity before proceedings are finalized.

\subsection{Multi-Objective Evaluation: Precision vs. Recall}
\begin{figure}[htbp]
    \centering
    \includegraphics[width=0.99\textwidth]{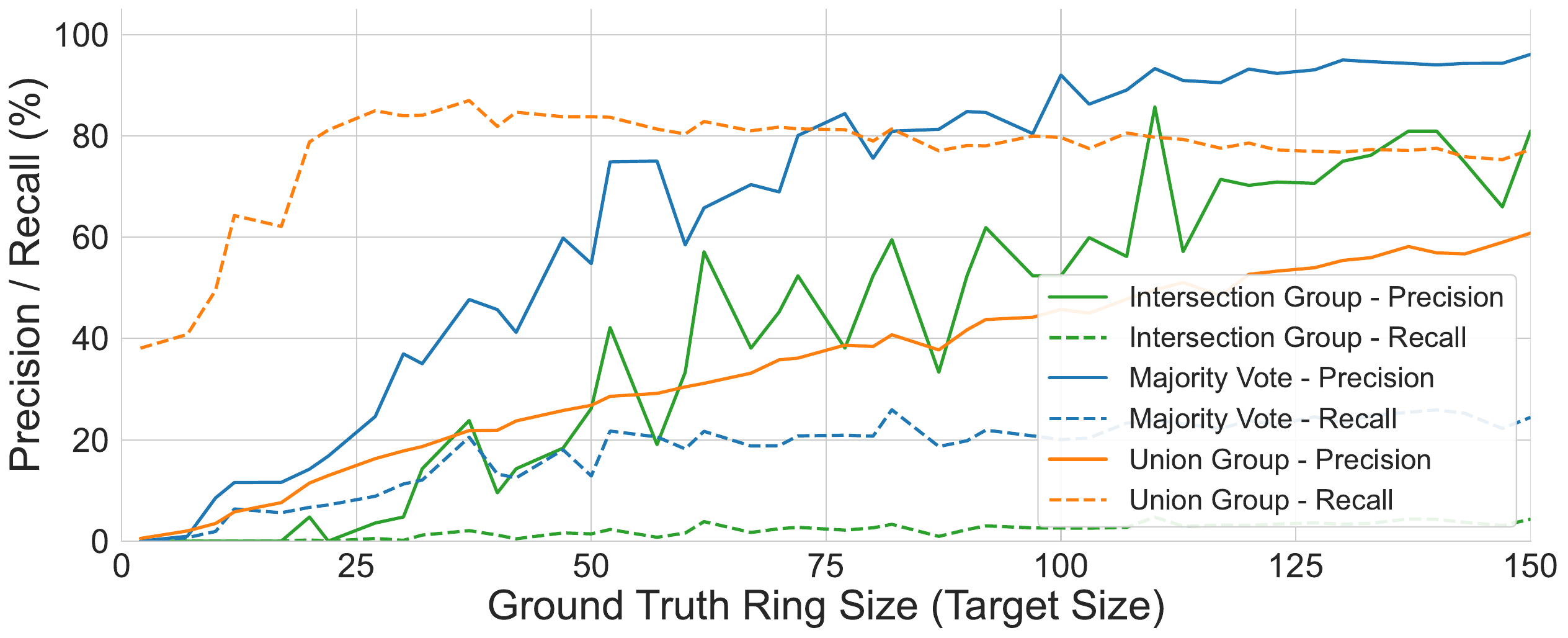}
    \caption{Precision and Recall dynamics. The Union Group optimizes for high Recall, while the Majority Vote optimizes for high Precision.}
    \label{fig:pr_dynamics}
\end{figure}

To formally evaluate the accuracy of the identified subsets, let $G^*$ denote the ground-truth collusion ring and $\hat{G}$ denote the flagged group. We measure Recall, defined as $\frac{|\hat{G} \cap G^*|}{|G^*|}$ (the proportion of the true ring detected), and Precision, defined as $\frac{|\hat{G} \cap G^*|}{|\hat{G}|}$ (the purity of the flagged group)\footnote{In standard classification terms, optimizing Recall minimizes False Negatives (missed colluders), while optimizing Precision minimizes False Positives (falsely flagging innocent reviewers).}. Figure~\ref{fig:pr_dynamics} demonstrates the inherent trade-offs among the consensus strategies defined in Section~\ref{subsec:iterative_refinement} ($G_{\text{inter}}, G_{\text{maj}}, G_{\text{union}}$).

The empirical results reveal two distinct operational modes. The \textbf{Union Group} prioritizes coverage, rapidly achieving a Recall of $\approx 80\%$ across all ground-truth sizes. However, this high sensitivity introduces topological noise, causing Precision to plateau at approximately 55\%.

Conversely, the \textbf{Majority Vote} prioritizes accuracy. It achieves high purity, with Precision reaching over 90\% for larger rings, indicating a very low false-positive rate. Meanwhile, its Recall remains conservative at roughly 20\%. The \textbf{Intersection Group} underperforms in this context; its minimal subset size severely bounds its Recall, and its Precision suffers from high variance, thereby reducing its viability as a reliable diagnostic tool compared to the Majority Vote.

\textbf{A Combined Approach for Program Chairs:} These contrasting dynamics provide a multi-tiered forensic tool for conference administrators. By initially deploying the \textbf{Union Group}, Program Chairs can isolate a highly suspicious sub-network that captures the vast majority ($>80\%$) of malicious actors. Within this bounded search space, applying the \textbf{Majority Vote} identifies the highest-confidence offenders with $>90\%$ Precision. This combined approach allows administrators to focus their auditing efforts effectively and make empirically informed decisions.

\subsection{Algorithmic Convergence and Statistical Power}
To complement the Precision-Recall analysis presented in Figure~\ref{fig:pr_dynamics}, we examine the sizes of the recovered groups and evaluate their statistical significance using the FWER controls described in Section~\ref{subsec:statistical_validation}.

\begin{figure}[htbp]
    \centering
    \subfloat[Discovered Size Tracking]{
        \includegraphics[width=0.46\textwidth]{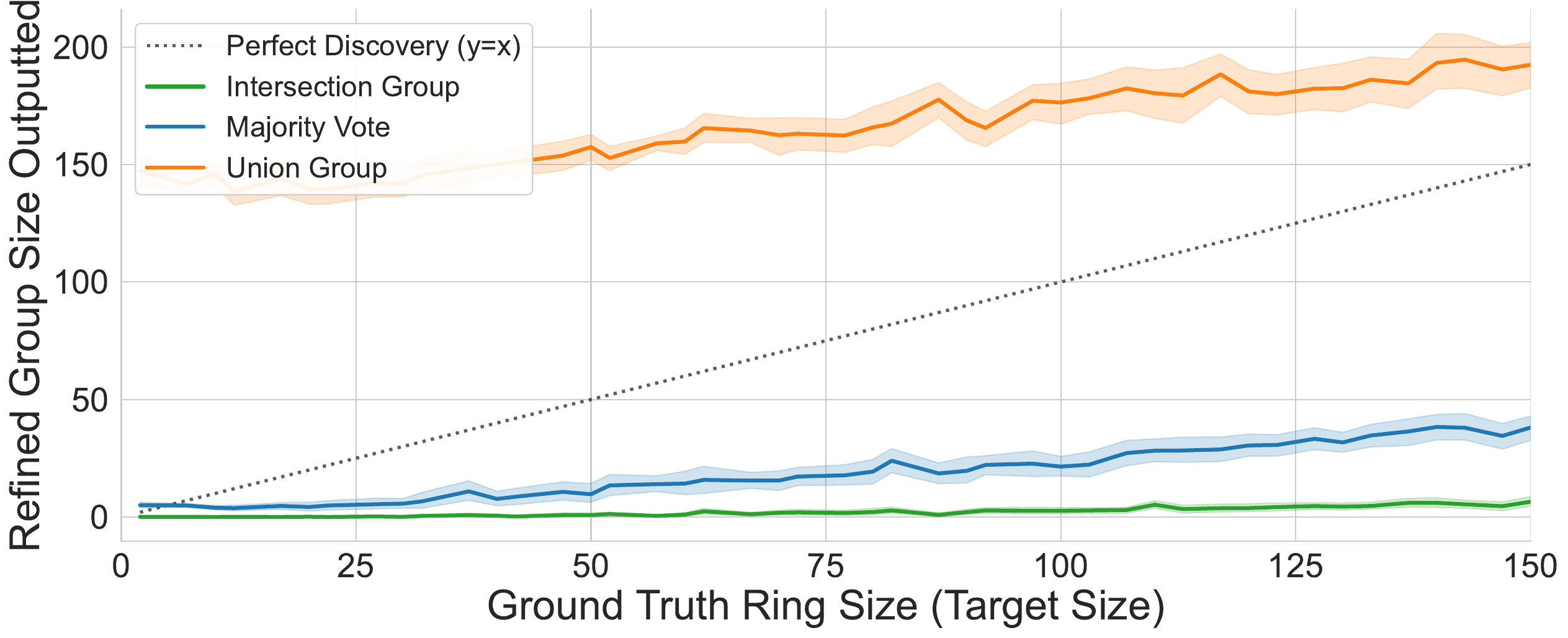}
        \label{fig:combined_size}
    }
    \hfill
    \subfloat[Statistical Power (FWER)]{
        \includegraphics[width=0.46\textwidth]{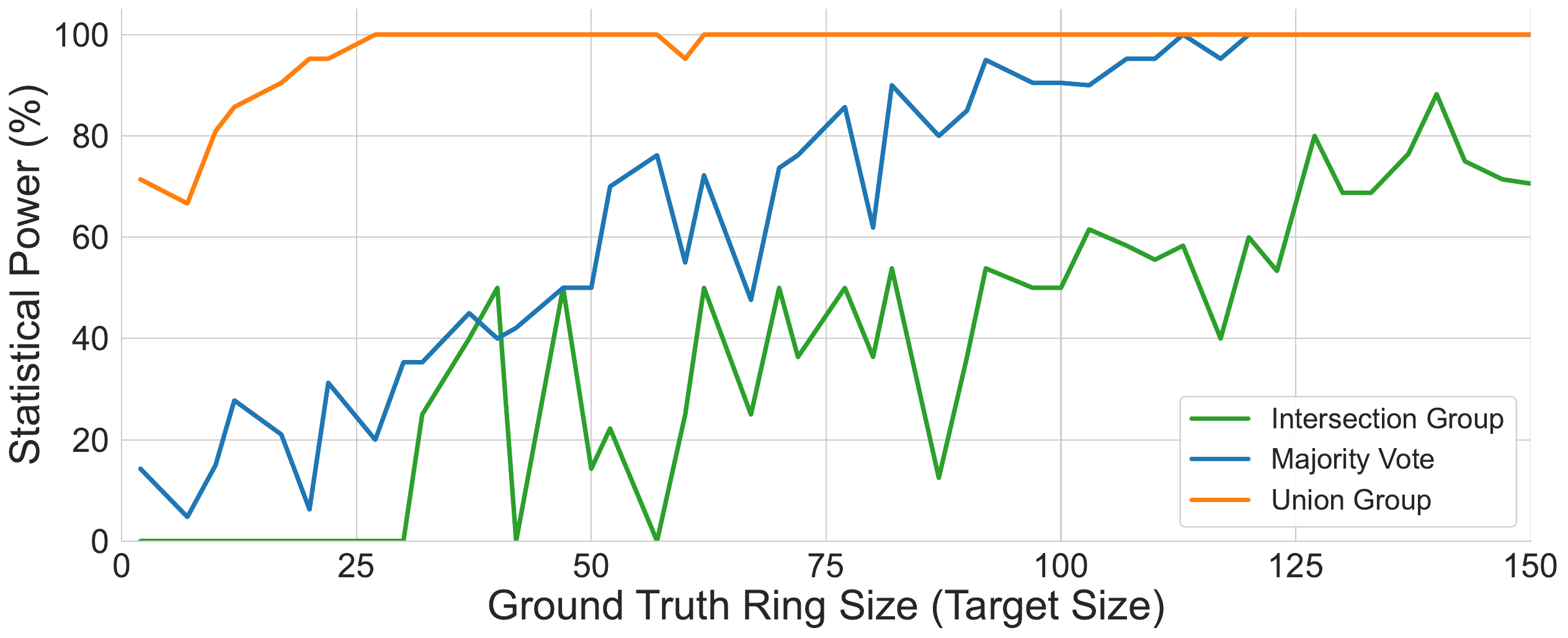}
        \label{fig:combined_fwer}
    }
    \caption{(a) Discovered group size versus ground-truth ring size. (b) Statistical Power under FWER control. The y-axis represents the percentage of iterations (out of 20 independent trials) where the identified group was deemed statistically significant.}
    \label{fig:convergence_fwer}
\end{figure}

Figure~\ref{fig:convergence_fwer}(a) illustrates the scaling behavior of the different consensus strategies. The \textbf{Union Group} consistently overestimates the suspected group size, capturing between 150 reviewers for small rings and approximately 200 for larger ones. In contrast, the \textbf{Majority Vote} scales proportionally with the injected ring, growing from single digits to approximately 50 reviewers. The \textbf{Intersection Group} remains highly restrictive, isolating a minimal subgraph of fewer than 15 reviewers regardless of the underlying manipulation scale. These observations contextualize the performance dynamics shown in Figure~\ref{fig:pr_dynamics}: The \textbf{Intersection Group} is simply too small to effectively encompass large collusion rings. The \textbf{Majority Vote} successfully scales with the ground-truth ring but remains conservative, capturing a highly pure subset of the actual colluders. Finally, while the \textbf{Union Group} returns a set significantly larger than the ground-truth ring, it scales proportionally and remains a localized subset relative to the entire conference population.

Figure~\ref{fig:convergence_fwer}(b) evaluates the statistical significance of the recovered groups across different injected ring sizes. The \textbf{Union Group} effectively isolates significant structural anomalies, achieving FWER-controlled significance in 100\% of the iterations for ground-truth rings as small as $|G| \approx 30$. In contrast, the \textbf{Majority Vote} and \textbf{Intersection Group} require larger injected rings to achieve statistical distinctness. This occurs because highly restrictive groups (fewer than 10 reviewers) produce a smaller overall effect on the review graph; therefore, it is mathematically plausible for a random group of the same size to achieve a similar anomaly score by chance (exceeding the $\gamma$ threshold). Consequently, the \textbf{Majority Vote} requires an injected size of $|G| \approx 120$ to consistently reach FWER control for 100\% of its iterations, while the highly restrictive \textbf{Intersection Group} peaks at merely a 60\% FWER rate, even at the maximum evaluated collusion size.

To further validate the framework's resistance to false discoveries, we executed the complete discovery pipeline on the ICLR baseline with no injected collusion. As anticipated, the independent algorithms returned exceptionally small, disjoint subsets with minor overlap. Consequently, the consensus formations yielded non-significant FWER scores and precision, empirically confirming the system's structural immunity to false alarms in an honest review environment. Evaluating subtle collusion strategies (Section~\ref{subsec:collusive_behavior}) showed similar precision-recall trends, though with proportionally lower FWER significance due to their smaller systemic footprint.

Operationally, these dynamics translate into a clear, three-step auditing protocol for conference organizers. First, the FWER-controlled significance flag ($\text{Is\_Sig}_G$) serves as the definitive trigger, strictly confirming the presence of an anomaly and mathematically justifying intervention. Once an alarm is validated, chairs can deploy the consensus formations strategically: the \textbf{Majority Vote} group acts as a high-confidence core for immediate scrutiny, while the broader \textbf{Union Group} serves as an inclusive watchlist to capture peripheral actors. This decoupled approach empowers organizers to move from purely observing statistical noise to executing confident, empirically informed policy decisions.

\section{Multiple Concurrent Collusion Rings}

\begin{figure*}[!ht]
    \subfloat[]{
        \includegraphics[width=0.46\textwidth]{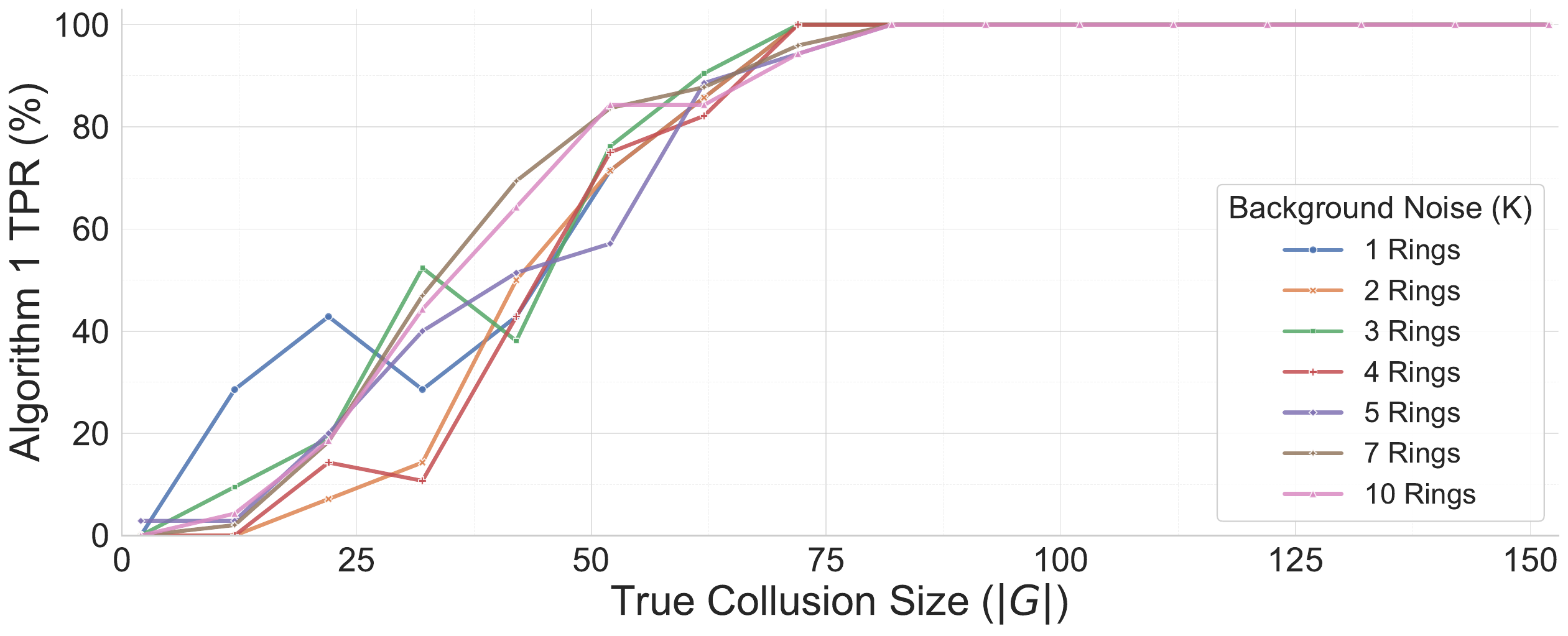}
    }\hfill
    \subfloat[]{
        \includegraphics[width=0.46\textwidth]{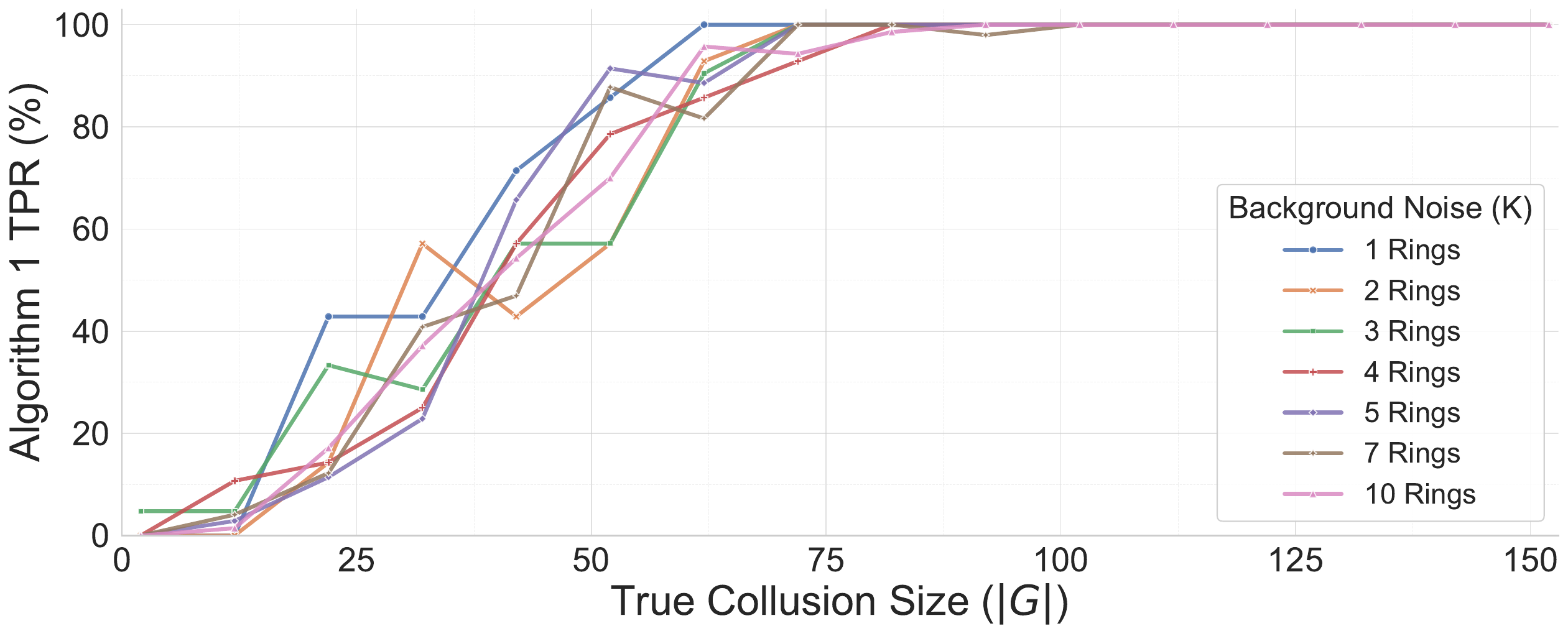}
    }
    
    \subfloat[]{
        \includegraphics[width=0.46\textwidth]{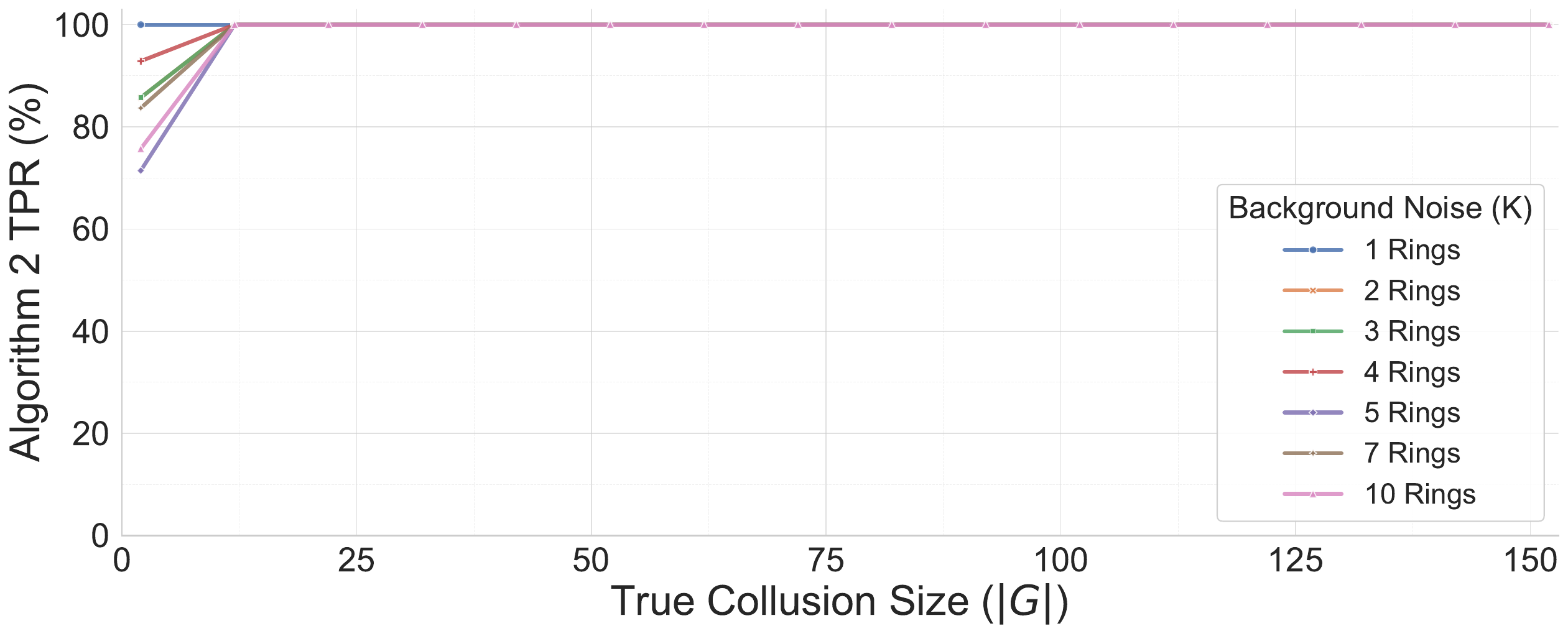}
    }\hfill
    \subfloat[]{
        \includegraphics[width=0.46\textwidth]{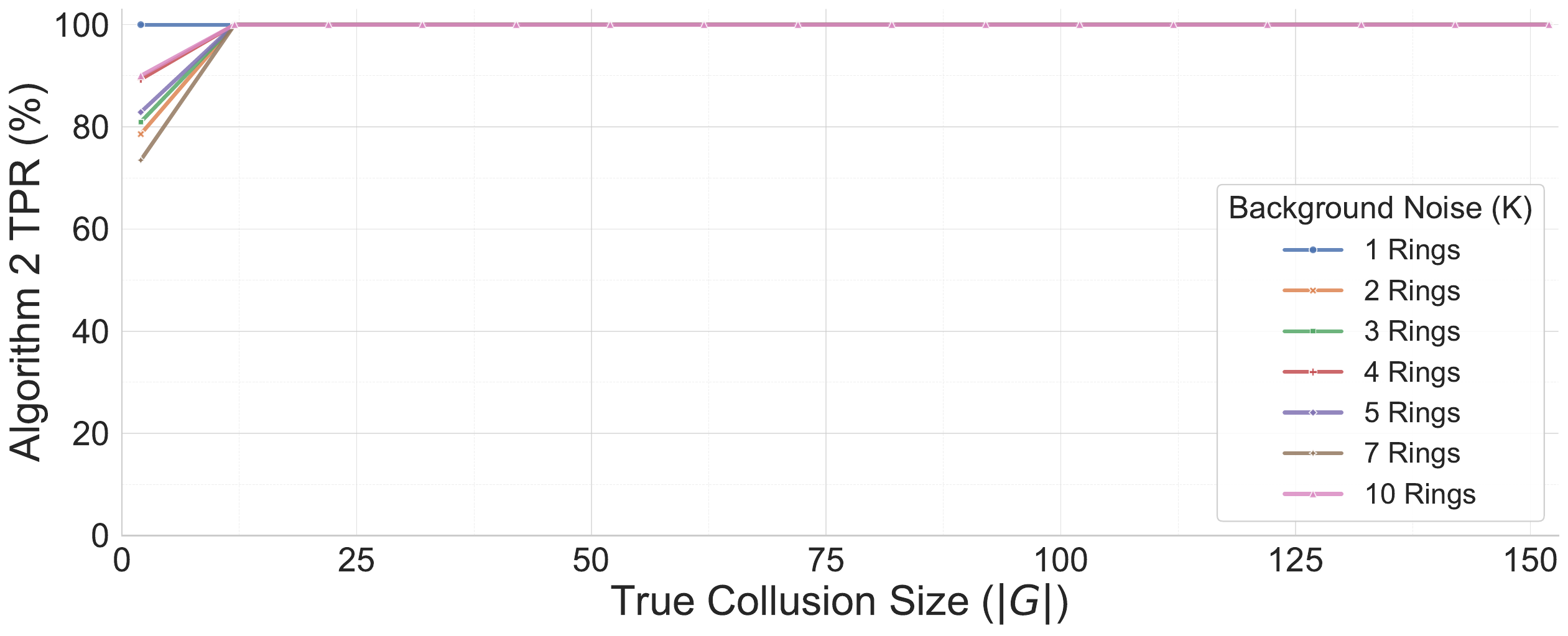}
    }
    
    \subfloat[]{
        \includegraphics[width=0.46\textwidth]{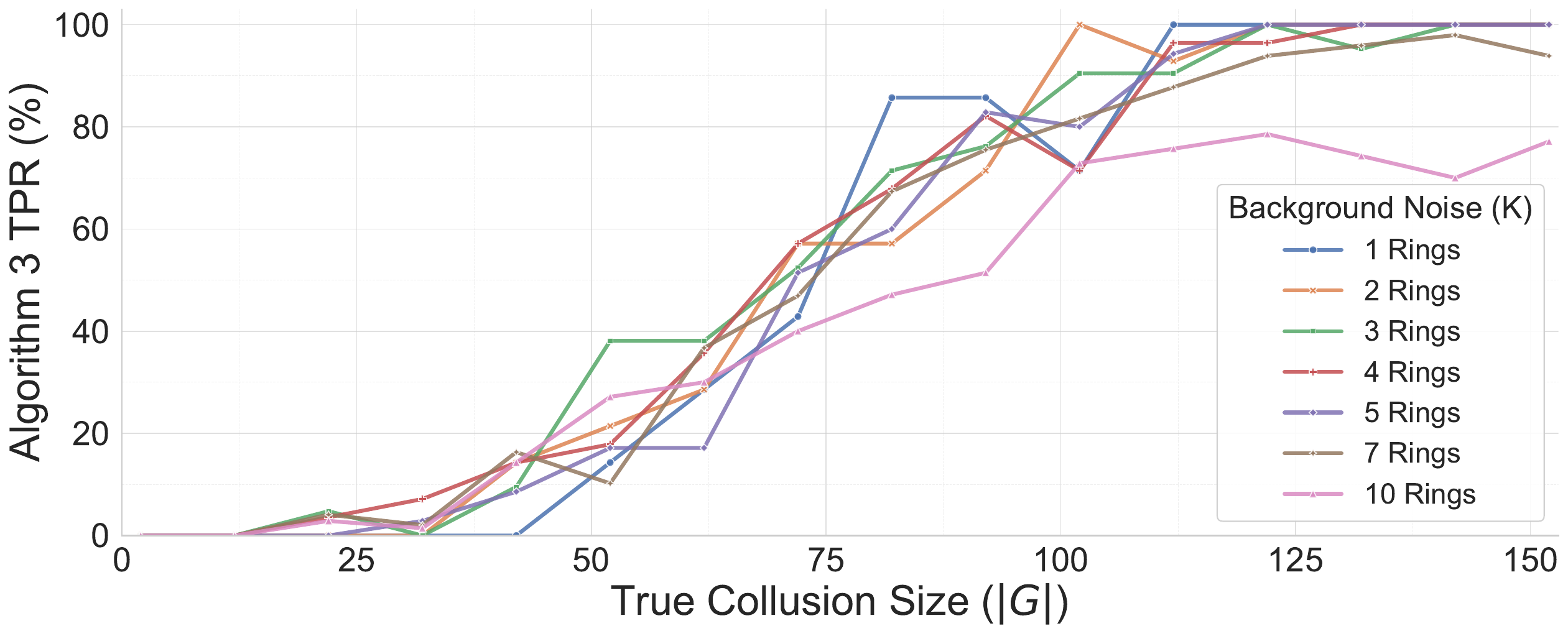}
    }\hfill
    \subfloat[]{
        \includegraphics[width=0.46\textwidth]{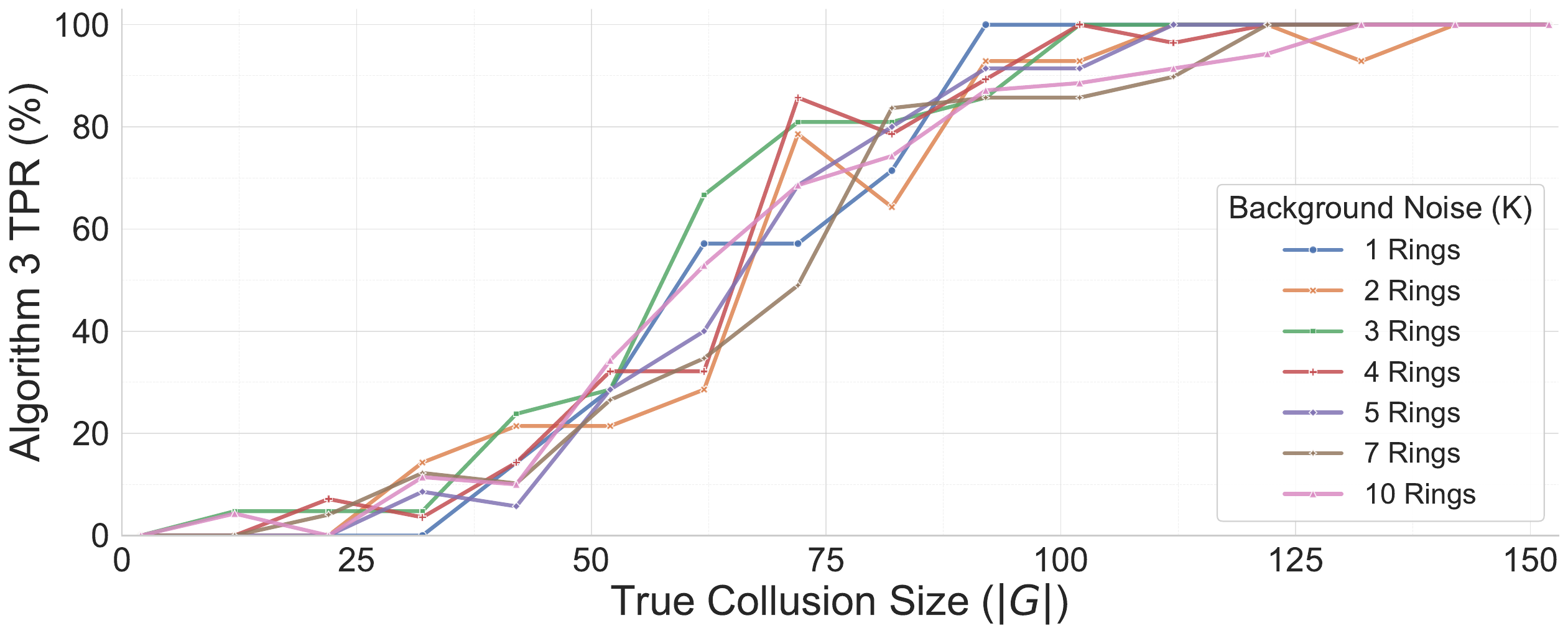}
    }
    
    \subfloat[]{
        \includegraphics[width=0.46\textwidth]{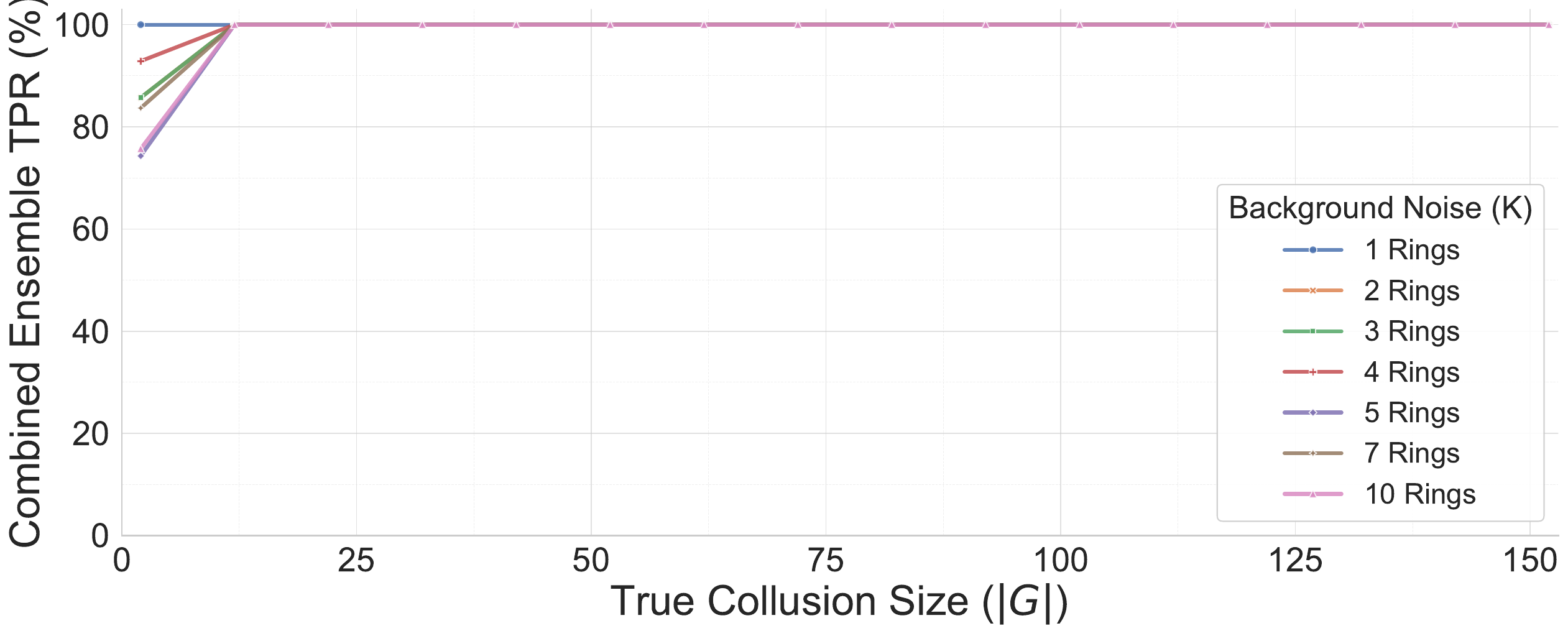}
    }\hfill
    \subfloat[]{
        \includegraphics[width=0.46\textwidth]{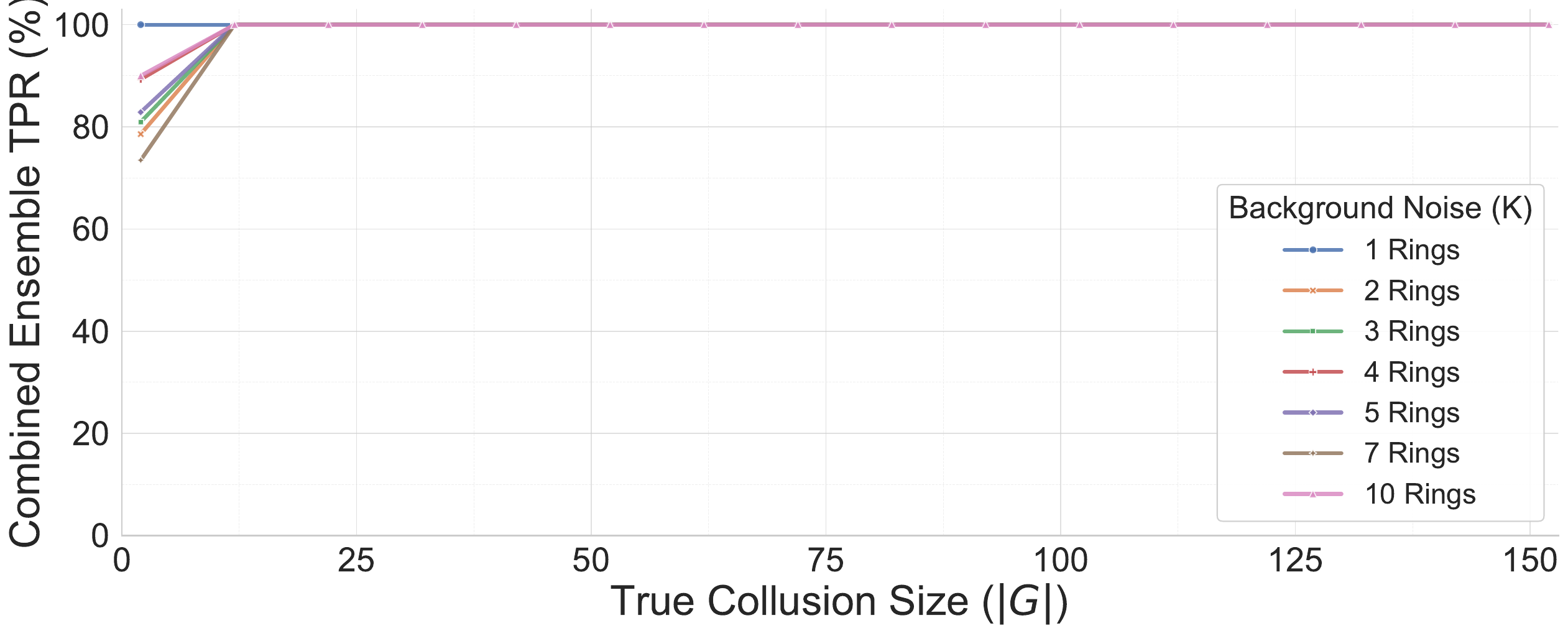}
    }
    
    \caption{Algorithm detection performance under multiple concurrent collusion rings, comparing disjoint (left) and overlapping (right) topologies. Rows (top to bottom): Algorithm 1, 2, 3 and combined.}
    \label{fig:multi_effect_tpr}
\end{figure*}

In large-scale conferences, multiple collusion rings may operate simultaneously. These groups can be disjoint, operating entirely independently with no shared reviewers, or overlapping, where specific actors participate in multiple rings, coordinating across different manipulative efforts. We extend our evaluation to verify the robustness of our framework under these complex multi-ring conditions.

\subsection{Multi Detection Algorithm Simulation}

To evaluate the core detection algorithms, we replicate the controlled simulation methodology from Section~\ref{sec:simulation_results}. We evaluated scenarios with 2, 5, and 10 concurrent rings, spanning from isolated coordination to widespread systemic saturation, utilizing a single injected ring as a control baseline. For all multi-ring experiments in this section, the injected coalitions exclusively employed the Fully Coordinated collusion strategy (Type 2, detailed in Section~\ref{subsec:collusive_behavior}). Across all configurations, the injected rings were of identical sizes, scaling from 2 to 150 members in increments of 10. All other environmental constraints, including the assignment bias and the permutation testing configuration, remained strictly identical to our initial single-ring evaluation.

As shown in Figure~\ref{fig:multi_effect_tpr}, the algorithms successfully capture structural anomalies across all manipulation behaviors. Detection performance remains highly stable, with no significant statistical difference observed between overlapping and disjoint ring structures. A degradation in detection capability occurs only under extreme systemic saturation, specifically when injecting 10 rings of 150 members each. At this scale, the majority of the reviewer pool becomes complicit, causing the collusion to dictate the baseline venue variance rather than deviating from it.

\subsection{Multi Discovery Framework Evaluation}

\begin{figure}[!ht]
    \centering
    \begin{minipage}{0.47\textwidth}
        \centering
        \includegraphics[width=\linewidth]{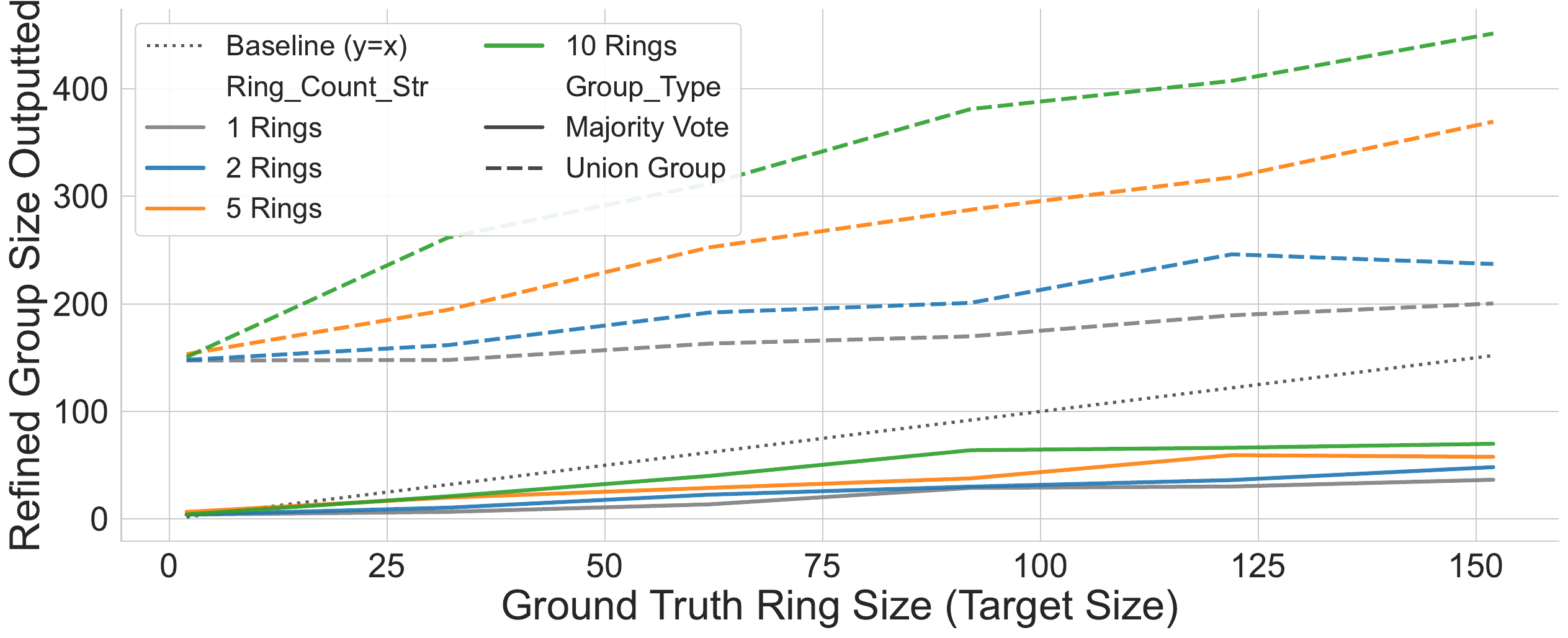}
        \par (a) Refined Group Size Dynamics
    \end{minipage}\hfill
    \begin{minipage}{0.47\textwidth}
        \centering
        \includegraphics[width=\linewidth]{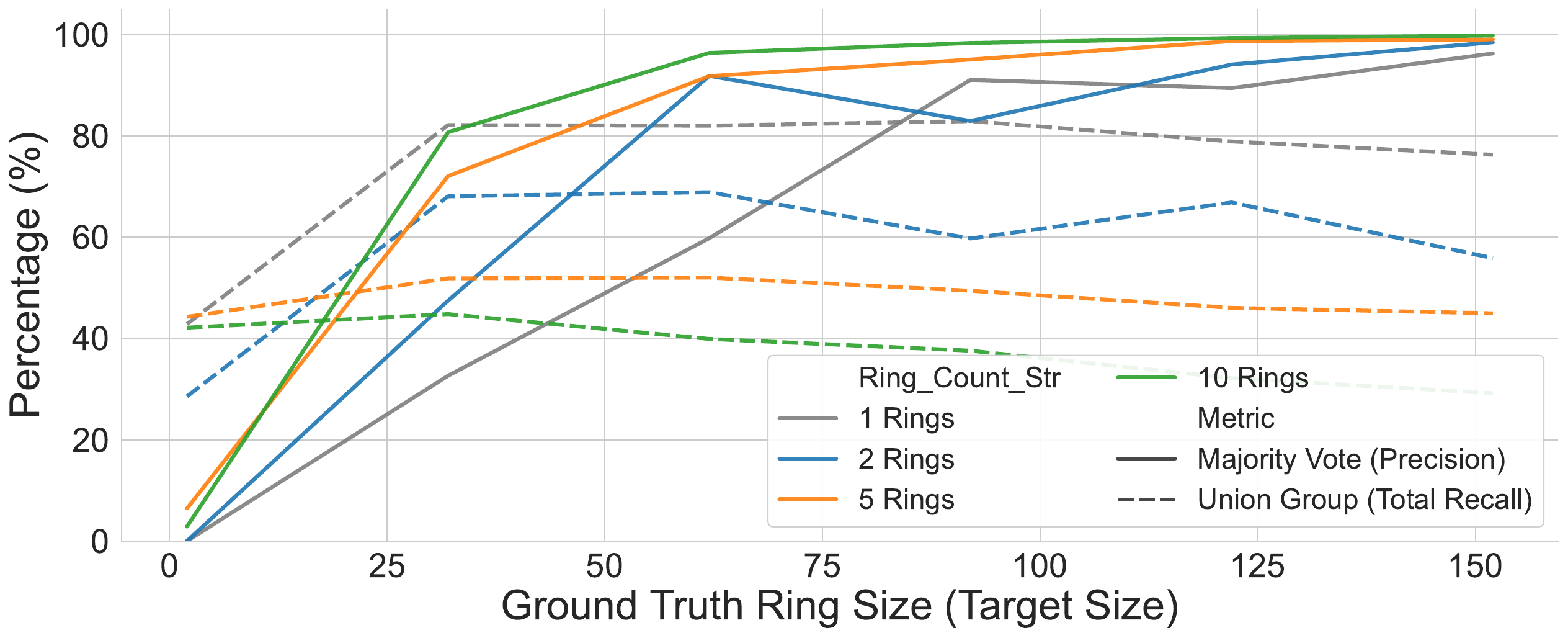}
        \par (b) Majority Vote Precision
    \end{minipage}
    
    \caption{Discovery engine metrics under multiple disjoint collusion rings, varying ring amounts ($K$) and sizes ($|G|$).}
    \label{fig:multi_discovery_metrics}
\end{figure}

Again, we executed our discovery process by embedding multiple rings into the ICLR semantic space, applying the exact same Gaussian noise injection and clustering initialization, and all properties remained as in the single ring case. The multi-group setting requires a careful definition of precision and recall. We define precision as the percentage of flagged reviewers who belong to any of the injected rings, and recall as the percentage of all colluding reviewers caught across the entire process.

As illustrated in Figure~\ref{fig:multi_discovery_metrics}, the size of the flagged groups grows as more rings are injected, directly reflecting the higher absolute number of colluding actors. Crucially, this growth remains proportional and within the same order of magnitude. Precision remains consistently high and marginally improves as the overall density of colluding reviewers increases. Conversely, Global recall exhibits a slight degradation.

The framework's ability to successfully isolate rings in a highly saturated environment counters the intuitive assumption that multiple uncoordinated attacks would generate enough noise to mask individual coalitions. This robustness can be attributed to two factors: each disjoint ring acts as a manipulating agent that we compute its marginal contribution. Multiple rings act as a set of independent manipulators that happen to coincide in the mechanism, but their effect remains separate.

\begin{figure}[!ht]
    \centering
    \subfloat[Refined Group Size]{
        \includegraphics[width=0.46\textwidth]{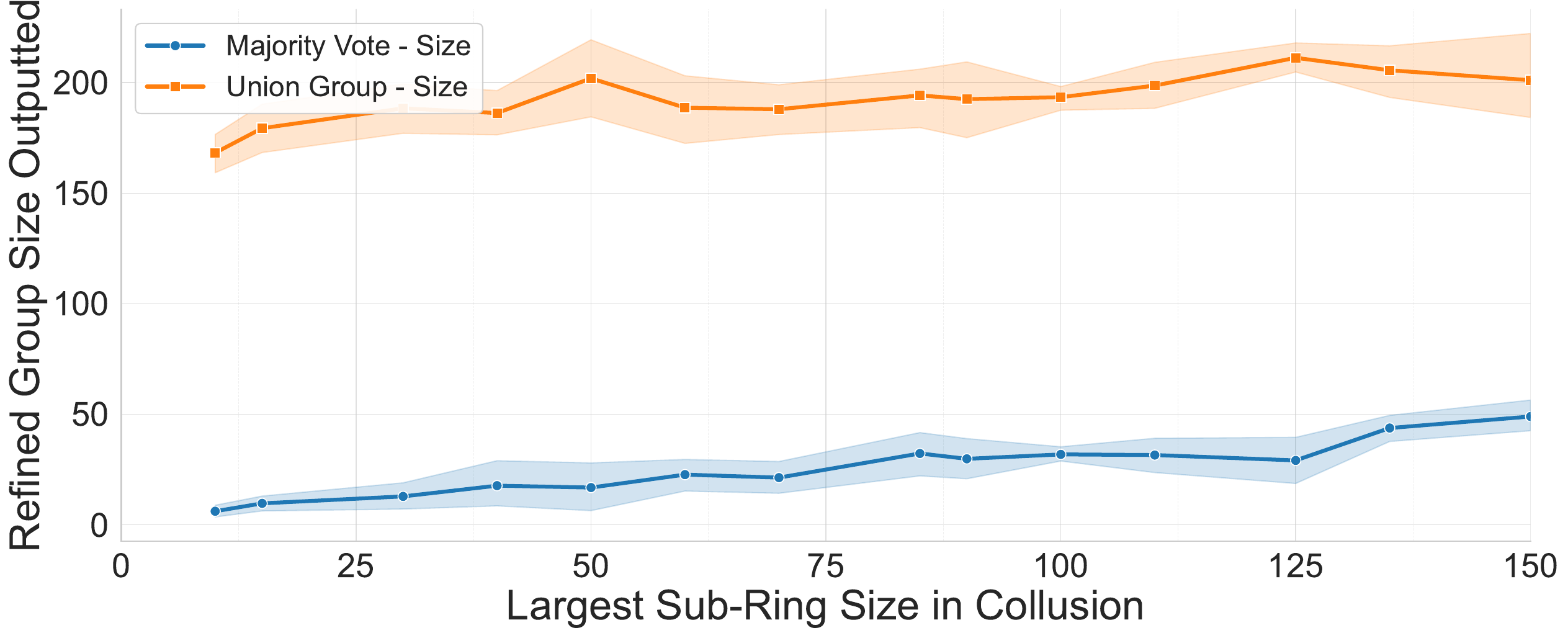}
    }
    \hfill
    \subfloat[Total Recall]{
        \includegraphics[width=0.46\textwidth]{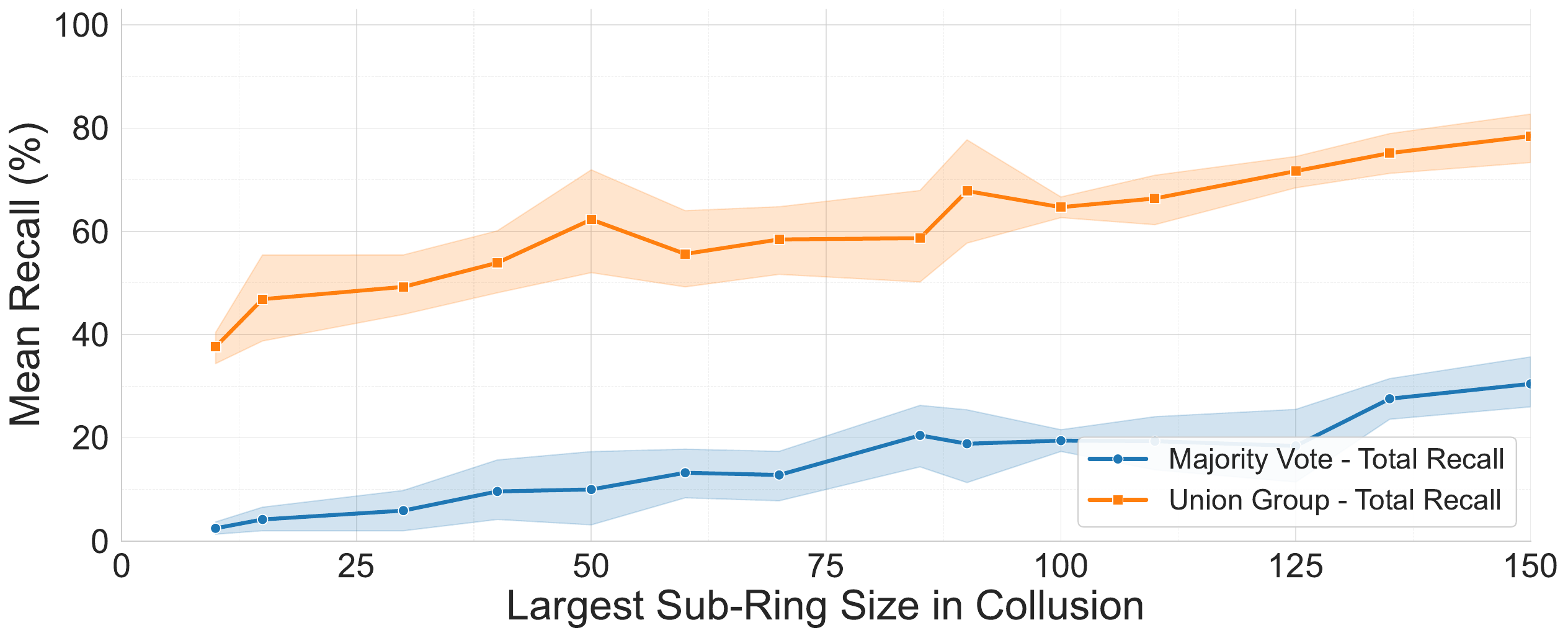}
    }
    \hfill
    \subfloat[Precision]{
        \includegraphics[width=0.46\textwidth]{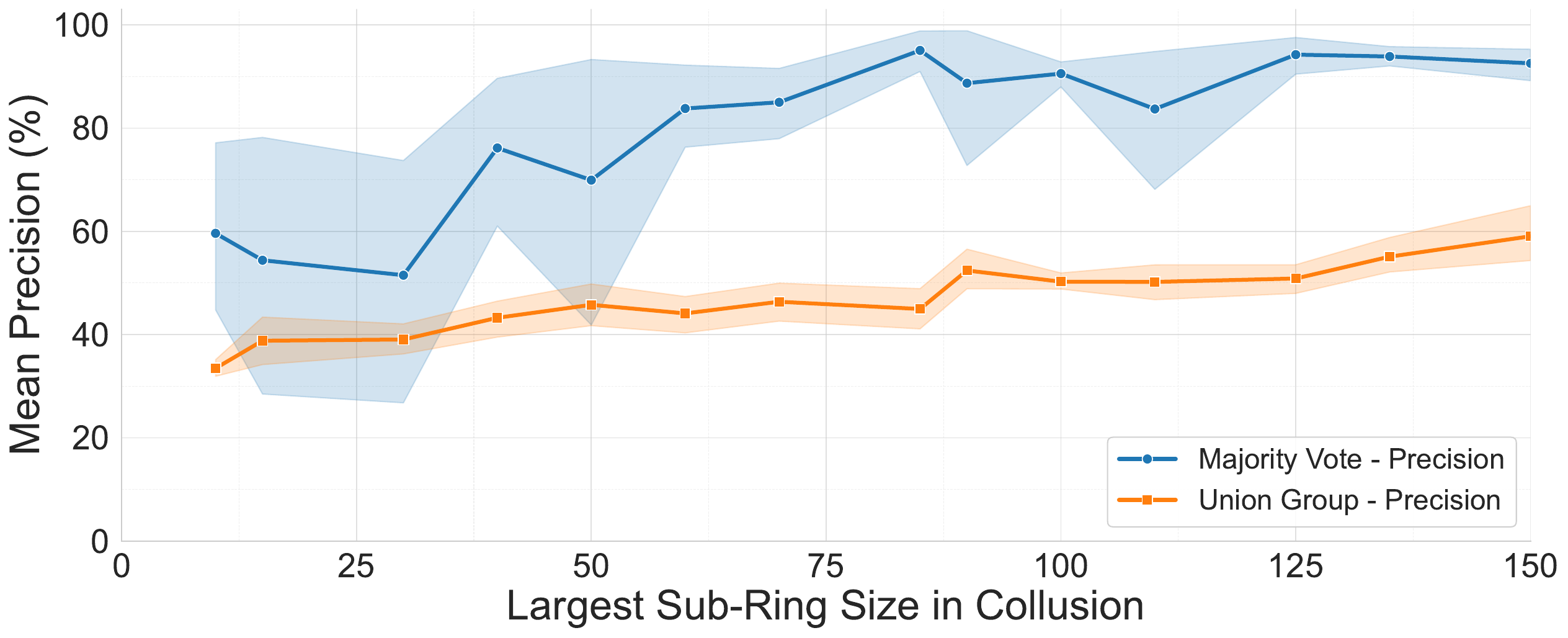}
    }
    
    \caption{Discovery engine performance under heterogeneous multi-ring collusion. Total colluders fixed at 150.}
    \label{fig:hetero_discovery_metrics}
\end{figure}

Figure~\ref{fig:hetero_discovery_metrics} evaluates heterogeneous fragmentation, fixing the total colluding population at 150 reviewers but splitting them into sub-rings of varying sizes. Crucially, the discovery mechanism's performance is driven more by the size of the largest individual ring, rather than the absolute number of colluding actors in the system. Highly fragmented networks (max sub-ring $<50$) impair the recall of the strict \textbf{Majority Vote} due to dispersed statistical anomalies. However, the \textbf{Union Group} acts as a robust safety net, maintaining a total recall between $40\%-80\%$ regardless of fragmentation, while the \textbf{Majority Vote} retains exceptional precision ($>70\%$) for the localized core it successfully identifies.

Ultimately, these findings confirm that the proposed detection algorithms and the discovery framework are structurally robust, effectively isolating manipulation even when confronted with multiple concurrent collusion networks.

\section{Discussion and Future Work}
In this paper we suggest two mechanisms: one, based on marginal-outcome changes, to examine if a group is a collusion ring, and another to examine a set of agents and try to find collusion rings in them. Our suggested systems do not rely on the specific design of the peer-selection algorithm used, and seem to be able to identify both overt and subtle collusion rings without prior knowledge of group membership, and work surprisingly well also in a multi-ring setting. 

This framework provides program chairs with immediate, actionable preventative measures. For the current review cycle, the high-precision \textbf{Majority Vote} group enables targeted editorial scrutiny, ensuring suspected members are excluded from evaluating associated borderline submissions. For subsequent venues, the broader \textbf{Union Group} serves as an inclusive watchlist. Organizers can deploy ``soft quarantines'' by dynamically expanding internal Conflict of Interest constraints, preventing flagged individuals from being assigned to each other and reducing their overall review load. Furthermore, aggregating historical matrices across multiple conferences into a longitudinal evaluation space will substantially increase statistical detection power.
Ultimately, by increasing the complexity and risk required to sustain coordinated manipulation, this framework progressively mitigates large-scale collusion rings, safeguarding the foundational trust of academic peer review.

Beyond hoping our proposed mechanism be used in actual conferences, this work suggests several enticing directions for future research: Can this approach be combined with intra-mechanism incentives to deter collusion (e.g., if one is suspected, their own paper is removed)? Can such mechanisms work better than current wack-a-mole systems?  How much information is sufficient to create a good embedding, and do email addresses and keywords suffice? We leave these questions to future research, building on our results.

\section*{Code and Data Availability}
To ensure full reproducibility, a ZIP archive containing the complete Python codebase and the ICLR and DPR datasets accompanies this manuscript. The repository provides a unified \texttt{main.py} entry point with a command-line interface for executing the documented experiments, alongside the \texttt{injector.py} module used for adversarial data generation. All required libraries and their specific version dependencies are detailed within the source code to guarantee a consistent runtime environment.

\section*{Acknowledgments}
We are grateful to Prof. Nihar Shah for his constructive feedback and insightful comments on an earlier version of this manuscript. We also extend our sincere thanks to Prof. Wolfgang E. Kerzendorf and his co-authors for publicly sharing the DeepThought Peer Review (DPR) dataset, which provided an invaluable foundation for our work.

\section*{Declaration of Competing Interest}
The authors have no competing interests to declare that are relevant to the content of this article.

\bibliographystyle{unsrtnat}
\bibliography{Full_Paper.bib}

\end{document}